\documentclass[longauth]{aa}
\usepackage{graphicx}
\usepackage{color}
\usepackage{xcolor}
\usepackage{soul}
\usepackage{hyperref}
\usepackage{amsmath}
\usepackage{longtable}
\usepackage[switch]{lineno}
\usepackage{txfonts}
\usepackage{subcaption}
\usepackage{multirow}

\begin{document} 
   \title{Selection and characterisation of the M-dwarf targets in the PLATO Input Catalogue}
    \titlerunning{III. characterisation of M-dwarfs targets in the PLATO Input Catalogue}
   \author{L. Prisinzano\thanks{E-mail: loredana.prisinzano@inaf.it}          \inst{\ref{ipris}}$^{\href{https://orcid.org/0000-0002-8893-2210}{\includegraphics[scale=0.5]{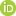}}}$
          \and
          M. Montalto
          \inst{\ref{imont}}
          \and
          G. Piotto
          \inst{\ref{ipiot1},\ref{ipiot2}}
          \and
           P. M. Marrese \inst{\ref{imarr1},\ref{imarr2}}
           \and
           S. Marinoni \inst{\ref{imarr1},\ref{imarr2}}
         \and
          V. Nascimbeni \inst{\ref{ipiot2},\ref{igran}}
           \and 
            V. Granata \inst{\ref{igran},\ref{ipiot2}}
           \and
          J.~Cabrera \inst{\ref{{icabr}}}
        \and
          K.~Belkacem \inst{\ref{igoup}} 
          \and
           M.~Deleuil \inst{\ref{idele}}
           \and
         L.~Gizon  \inst{\ref{igizo}} 
            \and
             M.~J.~Goupil \inst{\ref{igoup}} 
            \and  
           I.~Pagano \inst{\ref{imont}}
           \and
           D.~Pollacco \inst{\ref{ipoll}}
           \and
            R. Ragazzoni \inst{\ref{ipiot2}}
           \and
           H.~Rauer \inst{\ref{{icabr}},\ref{iraue}}
           \and    
   S.~Udry \inst{\ref{iudry}}
                 \and
          J. Maldonado\inst{\ref{ipris}}
           \and
           G. Micela \inst{\ref{ipris}}
           \and
           F. Damiani \inst{\ref{ipris}}
          \and
          L. Affer \inst{\ref{ipris}}
          \and
          G. Altavilla \inst{\ref{imarr1},\ref{imarr2}}
          \and 
          C. Argiroffi \inst{\ref{iargi},\ref{ipris}}
          \and
          S. Benatti \inst{\ref{ipris}}
          \and
          S. Cassisi \inst{\ref{icass}}
          \and
          R. Claudi \inst{\ref{ipiot2},\ref{iclau}}
           \and 
           S. Desidera \inst{\ref{ipiot2}}
            \and
            M. Fabrizio \inst{\ref{imarr1},\ref{imarr2}}
            \and
            E. Flaccomio \inst{\ref{ipris}}
            \and
            U. Heiter \inst{\ref{iheit}}
            \and
            A. F.~Lanza\inst{\ref{imont}}
            \and
            A. ~Maggio \inst{\ref{ipris}}
            \and
           L.~Malavolta \inst{\ref{ipiot1},\ref{ipiot2}}
           \and
           D.~Nardiello \inst{\ref{ipiot1},\ref{ipiot2}}
           \and 
           S.~Ortolani \inst{\ref{ipiot1},\ref{ipiot2}}
           \and
           A.~Sozzetti
           \inst{\ref{isozz}}
          }
   \institute{INAF-Osservatorio Astronomico di Palermo, Piazza del Parlamento 1, 90134, Palermo,Italy\label{ipris}
         \and
         INAF – Osservatorio Astrofisico di Catania, Via S. Sofia 78, 95123 Catania, Italy\label{imont}
           \and
           Dipartimento di Fisica e Astronomia "Galileo Galilei", Universit\`a degli Studi di Padova, Vicolo dell’Osservatorio 3, 35122 Padova, Italy\label{ipiot1}
           \and
           Centro di Ateneo di Studi e Attivit\`a Spaziali "Giuseppe Colombo", Universit\`a degli Studi di Padova, Via Venezia 1, 35131 Padova, Italy\label{igran}
             \and
             INAF – Osservatorio Astronomico di Padova, vicolo dell’Osservatorio 5, 35122 Padova, Italy  \label{ipiot2}
             \and
             INAF – Osservatorio Astronomico di Roma, Via Frascati, 33, 00078
              Monte Porzio Catone (RM), Italy\label{imarr1}
            \and
            SSDC-ASI, Via del Politecnico, snc, 00133 Roma, Italy\label{imarr2}
            \and
                       Institute of Astronomy, KU Leuven, Celestijnenlaan 200D, 3001, Leuven, Belgium \label{iaert}         
           \and       
            Dip. di Fisica e Chimica, Universit\`a di Palermo, Piazza del Parlamento 1, 90134 Palermo, Italy\label{iargi}
            \and 
Deutsches Zentrum für Luft- und Raumfahrt (DLR), Institut für Optische Sensorsysteme,\\
Rutherfordstraße 2, 12489 Berlin-Adlershof, Germany
\label{{iborn}}
\and
           Deutsches Zentrum f\"ur Luft- und Raumfahrt (DLR), Institut f\"ur Planetenforschung, Rutherfordstra{\ss}e 2, 12489 Berlin-Adlershof, Germany \label{{icabr}}    
\and
           INAF-Osservatorio Astronomico d’Abruzzo, via M. Maggini, sn. 64100, Teramo, Italy\label{icass} 
            \and
            Dipartimento di Matematica e Fisica, Universit\`a Roma Tre, Via
della Vasca Navale 84, 00146 Roma, Italy\label{iclau}  
           \and
           Aix-Marseille Universit\'e, CNRS, CNES, Laboratoire d’Astrophysique de Marseille, Technop\^{o}le de Marseille-Etoile, 38, rue Fr\'ed\'eric Joliot-Curie, 13388 Marseille cedex 13, France \label{idele}          
           \and
           Max-Planck-Institut f\"ur Sonnensystemforschung, Justus-von-Liebig-Weg~3, 37077~G\"ottingen, Germany \label{igizo}
           \and          
           LESIA, CNRS UMR 8109, Universit\'e Pierre et Marie Curie, Universit\'e Denis Diderot, Observatoire de Paris, 92195 Meudon, France\label{igoup}
           \and           
           European Space Agency (ESA), European Space Research and Technology Centre (ESTEC), Keplerlaan 1, 2201 AZ Noordwijk, The Netherlands \label{ihera} 
           \and          
            Centro de Astrobiolog\'{\i}a (CSIC--INTA), Depto. de Astrof\'{\i}sica, 28692 Villanueva de la Ca\~nada, Madrid, Spain \label{imas}         
            \and
            Center for Space and Habitability, University of Bern, Bern, Switzerland. \label{iosbo}    
            \and
            Department of Physics, University of Warwick, Gibbet Hill Road, Coventry CV4 7AL, UK \label{ipoll}          
            \and
            Armagh Observatory \& Planetarium, College Hill, Armagh, BT61 9DG, UK \label{irams} 
            \and           
            Institute of Geological Sciences, Freie Universit\"at Berlin, Malteserstra{\ss}e 74-100, 12249 Berlin, Germany \label{iraue} 
\and
Observational Astrophysics, Department of Physics and Astronomy,
Uppsala University, Box 516, SE-751 20 Uppsala, Sweden \label{iheit} 
     \and 
    INAF-Osservatorio Astrofisico di Torino, 
    via Osservatorio 20, 10025 Pino Torinese, Italy\label{isozz}   
   \and
   Observatoire de Gen\`eve, Universit\'e de Gen\`eve, Chemin Pegasi 51, 1290 Sauverny, Switzerland \label{iudry}        
             }
   \date{Received XX; accepted YY}

  \abstract
   {The European Space Agency's PLAnetary Transits and Oscillations of Stars 
   (PLATO) mission aims to detect  planets orbiting around 
   dwarfs and subgiant stars with spectral type F5 or later, 
   including M-dwarfs.
The PLATO Input Catalogue (PIC) contains all PLATO targets available for observation by the PLATO nominal science. 
The latest version, PIC\,2.1.0.1, focuses on the 
Southern PLATO field, named LOPS2, 
selected as the first long observation field,
and includes the P4 sample, one of the four target samples outlined in the Science Requirement Document. 
P4 includes   
 the M-dwarfs 
 with magnitudes brighter than V=16 
 located within LOPS2. }
   {A characterisation of the M-dwarfs 
   in the PIC
   is essential for
assessing their 
potentiality 
to host  exoplanets,
and eventually estimate the hosted planet(s) properties. 
The purpose of this paper is to describe how we selected the P4 M-dwarf targets, and obtained their
fundamental parameters and properties.}
{ 
Measuring stellar parameters is a challenging task:  interferometry provides direct estimates of radii, whereas alternative approaches relying on theoretical assumptions are still affected by significant uncertainties.
In this work, we introduce the P4 sample and detail the methodologies, 
all based on photometric criteria, adopted for the measurement of their stellar parameters.}
   {Based on a statistical analysis of the P4 sample, we assess both the photometric and volume completeness, and classify the stellar populations according to their Galactic spatial-velocity components. The adopted stellar parameters are validated by comparison with independent methods from the literature used to estimate stellar radii.
 }
   {We conclude that the P4 sample is compliant with the PLATO science requirements. Being magnitude limited, its volume completeness decreases going towards distances larger than 30\,pc, where late-type targets are progressively less covered. The observed large spread in the colour-magnitude diagram is likely
   due to 
the combination of several effects such as metallicity, age, binarity and activity. The strategy we adopted for deriving stellar parameters provides results consistent with those obtained in the literature with different and independent methods. }


\maketitle
%
\section{Introduction \label{sec:intro}}
The PLAnetary Transits and Oscillations of Stars (PLATO) mission \citep{raue14,raue25} is an M-class mission of the European Space Agency (ESA).
It is scheduled to be launched in late 2026 and its primary aim is to search for and characterize terrestrial exoplanets  orbiting stars similar to our Sun   by employing the transit method. 
The PLATO mission will generate two primary types of data: light curves and imagettes
but, due to telemetry limitations, only those of pre-selected targets will be downloaded and analysed. 
The PLATO target samples will be 
located in the  Long-duration Observation Phase (LOP) PLATO fields \citep{nasc22,nasc25}, and are 
presently available to all PLATO consortium members
through  the PLATO Input Catalogue (PIC). 
A version of the PIC will become public nine months before PLATO launch. We refer to \citet{mont21}
 for a complete description of the PLATO samples.
 
PLATO's primary mission objectives 
focus on observing F, G, and K spectral type stars, including Sun analogues
and the targets are formally grouped into four  samples \citep{mont21}, in agreement with the ESA PLATO Science Requirements  \citep{raue25}.
One of these is the P4 sample, which 
contains 
cool late-type dwarf targets, that are the focus of this paper.

The M-type main sequence stars selected for the P4 sample  are of great interest in the field of extrasolar planets for several reasons.
The stellar initial mass function (IMF) peaks between 0.1 and 0.5\,M$_\odot$, 
making
M-type stars 
the most common stars, with 
a single star fraction reaching values of  $\sim 75$\% \citep{lada06}.
   In addition, M-type stars have a better star/planet luminosity contrast with respect to solar-type stars.
    Considering a planet with fixed mass and orbital properties, stellar radial velocity semi-amplitude due to the presence of a planet is a factor $\gtrsim 2$ larger 
    for M-dwarfs 
    than 
    for solar-type stars.
From the photometric point of view, the transit depth scales with the square of the
ratio between the planetary and stellar radii $R_P/R_\star$. This means that, for a fixed planetary size, smaller host stars produce deeper transits, making them significantly easier to detect \citep{nasc12,lovi17}.
Finally, the habitable zone (HZ) is closer  (within 0.1-0.2 AU) to the host star \citep{kopp13}
    and this increases the probability to detect transiting planets in 
    HZ by a factor 2-3, as well as the transit frequency during observations.
This makes such a sample particularly important for statistically probing the inner edge of the HZ, including the hypothesised HZ Inner Edge Discontinuity, as demonstrated in simulations computed for PLATO \citep{schl24}.

Planet formation around M-dwarfs is fundamentally challenged by their low-mass protoplanetary disks, which contain insufficient material for traditional core accretion to build gas giant cores before the gas dissipates. The leading solution is the pebble accretion model, where a growing core can efficiently sweep up pebble-sized particles, rapidly assembling the mass needed to form the super-Earths and mini-Neptunes commonly observed around these stars \citep{lamb12}. The alternative disk instability model is considered unviable, as M-dwarf disks lack the requisite mass to become gravitationally unstable \citep{krat16}. Consequently, the observed rarity of gas giants and abundance of smaller planets around M-dwarfs serves as a crucial observational test, strongly favouring pebble accretion as the dominant formation mechanism in low-mass stellar systems \citep{dres15}.

M dwarfs are intrinsically faint in the optical bands, and mostly emitting in the near-infrared (NIR). Moreover, they are strongly affected by magnetic activity 
\citep{koch21}
and their spectra are dominated by molecular bands, which complicates their characterisation, particularly the precise determination of stellar parameters and chemical abundances \citep{mald15, mald20, olan25}. Determining stellar radii is especially challenging due to their intrinsically small sizes, and only a limited number of direct radius measurements obtained through long-baseline interferometry are available \citep{boya13}. 
Direct radii can be also obtained from SB2 eclipsing binaries (EBs)  using dedicated NIR spectroscopy \citep{maxt22}. 
Nonetheless, a significant advantage is that M dwarfs are the most numerous stellar type in the Galaxy and, due to their high local spatial density, a large number of them are found in the solar neighbourhood. This makes them an ideal population for statistical studies and for testing stellar structure and evolution models through comparisons with direct measurements \citep{cass19}.

As first found by \citet{berg06},  direct  interferometrically derived radii are 15-20\% larger than those predicted 
by theoretical models based on bolometric luminosity.
Such a discrepancy, also known as "radius inflation" problem, has been 
 found also for detached, double-lined EBs
 for which precise and accurate measurements of stellar masses and radii have been determined \citep[e.g.][]{torr07,irwi11}. 
More recent interferometric and Gaia-based studies of single, slowly rotating, magnetically inactive M dwarfs indicate that radius inflation in single stars is generally less pronounced, typically in the 0–7\% range, compared to 5–20\% in eclipsing binaries 
\citep[e.g.][]{kess18,morr19, wand24, kima24}. This suggests that while binarity and rapid rotation play a significant role, other mechanisms such as magnetic activity, starspots, metallicity, and stellar age may also contribute.
\citet{berg06} found some evidence that the reported discrepancies correlate with metallicity, increasing with higher metal content. They interpreted this effect as evidence of shortcomings in the current generation of opacity tabulations used in stellar modelling.
Nevertheless, the presence of magnetic activity and, in particular, correlated starspots, has long been considered  \citep{spru82,spru86, macd12,cass13,feid13,feid14} the most plausible explanation for the radius inflation problem.
 Starspots reduce the effective radiating surface area of a star, leading to the radius inflation. 
 A treatment of the effects of starspots at the photospheric level on the stellar evolution models by \citet{togn18}, has been presented in  \citet{fran22} showing also that other stellar parameters, such as effective temperatures, are affected by the presence of starspots.
 
\citet{sway24}
measured mass, radius, and effective temperature of 23 M-dwarf companions to solar-type stars using CHEOPS observations. Using high-precision light curves from CHEOPS and comparing them with TESS data, for most targets they found evidence of  radius inflation.
The authors accounted for starspot-induced variations in their measurements and found trends linking radius inflation to metallicity.

Recent studies have shown that single, magnetically inactive M dwarfs exhibit smaller or negligible radius inflation, while active single stars can still display significant inflation, particularly at masses around 0.5–0.6 M$_\odot$ — exactly the stellar mass range targeted by PLATO. At lower masses, the effect appears less pronounced, which may reflect either larger measurement uncertainties or intrinsic physical differences in the stellar interior structure \citep[e.g.][]{kess18, wand24, kima24}.

In this work we present the selection and 
characterisation of the P4 sample included in the PIC\,2.1.0.1 located within the 
first LOP   
\citep[LOPS2,][]{nasc25}
footprint, in order to give a general astrophysical description of the sample.
The methods adopted for the sample selection and the parameter derivation are described in Sect.\,\ref{sec:methods}.
The results, including the photometric and volume completeness of the sample and the analysis of the kinematics are presented in Sect.\,\ref{sec:results}. A discussion about the comparison of the stellar radii with other studies and on the possible effects of the observed spread in the Gaia colour-absolute magnitude diagram  are presented in Sect.\,\ref{sec:discussion},
while our conclusions are given in Sect.\,\ref{sec:conclusions}.

\section{Methods\label{sec:methods}}
\subsection{Data  \label{sec:data} }
The main reference catalogue for the selection
of the targets of the most recent version of the PIC, including the P4 sample, is Gaia DR3 \citep{prus16,vall23}. 
As for the distances, we adopted the values provided by \citet{bail21}.
In addition,
to derive the stellar parameters of the P4 sample, we also used 2MASS data \citep{skru16}. 

The main strategy adopted to select the targets for the Plato samples, including the P4 sample
is described in \citet{mont21}, in which an all-sky version of the PIC (asPIC\,1.1.0), based on Gaia DR2 data, has been presented. 

A new version of the PIC, named PIC\,2.1.0.1, has been released to the PLATO Consortium on 21 February 2025. This version is limited to the LOPS2 footprint \citep[LOPS2]{nasc25} and is based on Gaia DR3. It also includes additional subPICs: fine guidance stars (fgPIC), instrument calibration targets (cPIC), and stellar astrophysics calibration targets (scvPIC).
 This paper is based on the PIC\,2.1.0.1.

\subsection{Stellar sample selection  \label{sec:selectionp4} }
\begin{figure}
	\centering
	\includegraphics[width=\columnwidth]{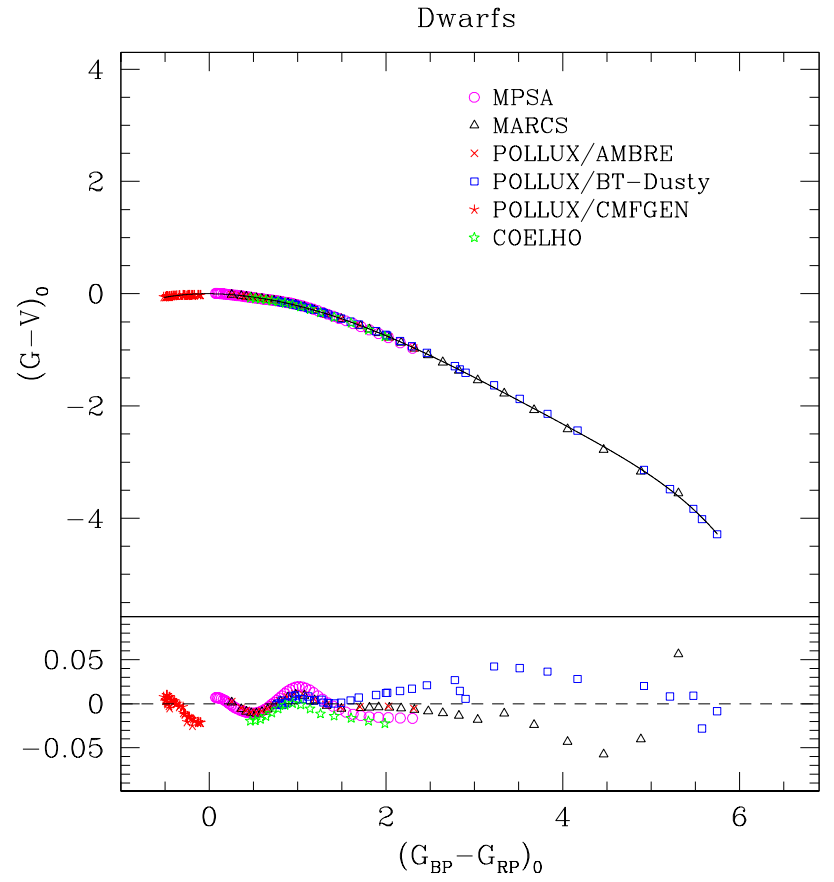}
	\caption{
Top panel: synthetic two-colour diagram of $(G - V)_0$ versus $(G_{\rm BP} - G_{\rm RP})_0$ for dwarf star models taken from the stellar libraries indicated in the legend. The solid line represents the best-fit polynomial to the synthetic colours.  
Bottom panel: residuals between the synthetic $(G - V)_0$ values and the best-fit relation, plotted as a function of the synthetic $(G_{\rm BP} - G_{\rm RP})_0$ colour.}
	\label{fig:GV_BPRP_DWARFS}
\end{figure}

The PLATO science requirements
impose 
the following conditions:
\begin{itemize}
    \item the total number of 
cool late-type dwarfs in stellar sample 4, P4, (cumulative over all sky fields
observed by PLATO) 
shall be at least 5,000, 
to be
monitored during a Long-Duration Observation Phase; 
    \item the dynamic range of stellar P4 shall be m$_V \lesssim$ 16;
    \item it shall be possible to obtain imagettes of 5,000 targets in stellar
sample 4 with a sampling time equal to 25 seconds. 
\end{itemize} 
Since the selection must be based on the visual $V$ magnitude,  this quantity was included in the PIC for each target.
The visible magnitude
was set equal to the Johnson $V$ magnitude when available in the literature; otherwise, it was 
estimated.
Since literature available photometric conversions  do not cover the low mass  M-type regime, we derived 
a photometric calibration providing a new conversion from the Gaia DR3 photometric system to the Johnson-Cousins system.
The calibration relation that we derived 
from  a least-square best-fit procedure is a sixth-order polynomial of the form:
\begin{equation}
\label{eq:pv0_bv0}
    (G-V)_0=\sum_{i=1}^{i=6} b_i(G_{BP}-G_{RP})_0^i,
\end{equation}
where the colours are those of dwarf star models taken from the MPSA, MARCS, POLLUX/AMBRE, POLLUX/BT-Dusty, POLLUX/CMFGEN, and COELHO stellar libraries.
The relation is valid in the range $-0.51 \le (G_{\rm BP}-G_{\rm RP})_0 \le 5.75$.
\noindent
The best-fit coefficients are listed in Table~\ref{tab:best_fit_coefficients_dwarfs_gv_bprp}, and the corresponding relation is illustrated in Fig.~\ref{fig:GV_BPRP_DWARFS},
showing the result, the adopted colours and the residuals of the fit.

\begin{table*}
\caption{Best-fit coefficients of the $(G-V)_0$ vs. $(G_{\rm BP}-G_{\rm RP})_0$ relation for dwarf stars. Validity range: $-0.51 \le (G_{\rm BP}-G_{\rm RP})_0 \le 5.75$.}
\label{tab:best_fit_coefficients_dwarfs_gv_bprp}
\centering
\begin{tabular}{ccccccc}
\hline\hline
$b_1$ & $b_2$ & $b_3$ & $b_4$ & $b_5$ & $b_6$ & $\sigma$ \\
\hline
$-0.0085337$ & $-0.2422638$ & $0.0493499$ & $-0.0161423$ & $0.0038450$ & $-0.0003356$ & 
0.017\\
\hline
\end{tabular}
\end{table*}
Given the very small uncertainties in the Gaia DR3 $(G_{\rm BP}-G_{\rm RP})_0$ colours for stars within the considered magnitude range, the low expected extinction, and the limited effect of the sixth-order polynomial (at most a few hundredths of a magnitude), the choice of a sixth-order fit does not impact the selection of the P4 sample.

The selection of the P4 sample targets has been performed by assuming the spectral type calibration given in \citet{peca13}
in which the spectral type M0 corresponds to T$_{\rm eff}$ = 3850\,K. To define the magnitude and colour boundaries for the selection in the colour-absolute magnitude diagram (CAMD), we adopted the Galactic simulations TRILEGALv1.6 \citep{gira05} falling in the 
region of the LOPS2 field.

 In order
to define the blue selection boundary, we considered two representative samples. The first
one
includes the targets, defined as stars having $T_{\rm eff} < 3850$ K and $\log g > 3.5$ and $V \leq 16$, as obtained from the Galactic simulations. These correspond to M dwarfs. The second sample includes the contaminants, defined as all stars with $V \leq 16$ that do not meet the target criteria, i.e., stars with $T_{\rm eff} > 3850$ K and/or $\log g < 3.5$. In this context, the contaminants are mainly FGK dwarfs. For the blue boundary, the goal is to separate M dwarfs (the 
P4
targets) from FGK dwarfs (the contaminants) in the colour–magnitude diagram. 
The best separation colour boundary was defined as the line around the blue limit of the targets, maximising the metric $S = (N_{\rm targ} - N_{\rm cont})$, where $N_{\rm targ}$ and $N_{\rm cont}$ are respectively the number of targets and contaminants.
The final separation boundary in this case corresponds to the theoretical best separation boundary and is given by the equation
${M}_{G,0} =-8.62(G_{\rm BP}-G_{\rm RP})_0 +24.96$, where $\rm M_{G,0}$ is absolute magnitude in the Gaia $G$ band, corrected for 
absorption
and  ($G_{\rm BP} - G_{\rm RP})_0$ are the reddening corrected Gaia colours.
Absorption and reddening were derived using the \citet{lall18} dust map, as described in \citet{mont21}.

As for the PIC\,1.1.0,
the boundary at the bright and red side of the M-dwarf selection region
 corresponds to the best regression line of a 10 Myr solar metallicity isochrone from the PARSEC database \citep{bres12}. The separation boundary is given by the following equation: $M_{G,0}=2.334(G_{\rm BP} -G_{\rm RP})_0+2.259$.

 In conclusion, the selection criteria adopted for the P4 sample are 
\begin{equation}
\label{eq:p4sel}
\left\{
\begin{aligned}
M_{G,0} &\geq 2.334(G_{BP} - G_{RP})_0 + 2.259 \\
M_{G,0} &> -8.62(G_{BP} - G_{RP})_0 + 24.96 \\
V &\leq 16.
\end{aligned}
\right.
\end{equation}

\begin{figure}
    \centering
\includegraphics[width=0.5\textwidth]{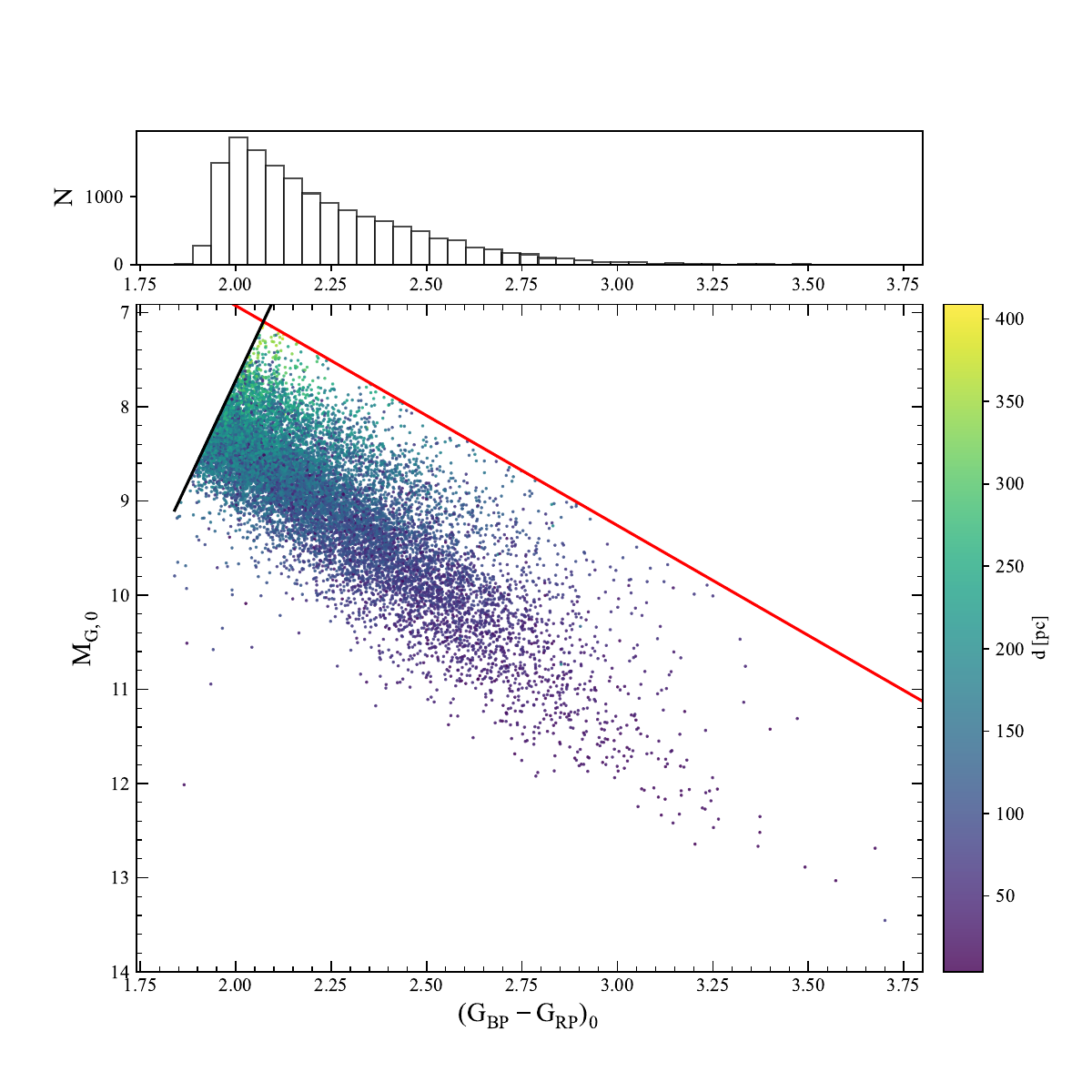}
    \caption{colour-absolute magnitude diagram of the P4 sample targets. The colour map indicates the distance of the targets. The black and red lines represent the adopted
    limits 
    indicated by the first two relations given in the Eq.\,\ref{eq:p4sel}.  The histogram of the $(G_{\rm BP}-G_{\rm RP})_0$ colours is also shown in the top panel.}
    \label{fig:p4mg0borp0}
\end{figure}

With these conditions we selected  
15140 targets 
suitable for the P4 sample.
Fig.\,\ref{fig:p4mg0borp0} shows the 
CAMD of the selected targets.
The figure illustrates also the  
distances
of the targets of the P4 sample,
and 
the histogram shows how the number of targets decreases towards later spectral types, which are intrinsically fainter and therefore detected only at shorter distances.

The M dwarf sample consists of 15157 targets obtained adding to the P4 sample, defined above, 17 additional M dwarfs which satisfy the first two conditions in  Eq.\,2, but have $V$>16.
The additional targets are M dwarfs with confirmed and/or candidate planets retrieved from the catalogue Exo-MerCat \citep{alei20,alei25}.

\begin{figure}
    \centering
\includegraphics[width=0.5\textwidth]{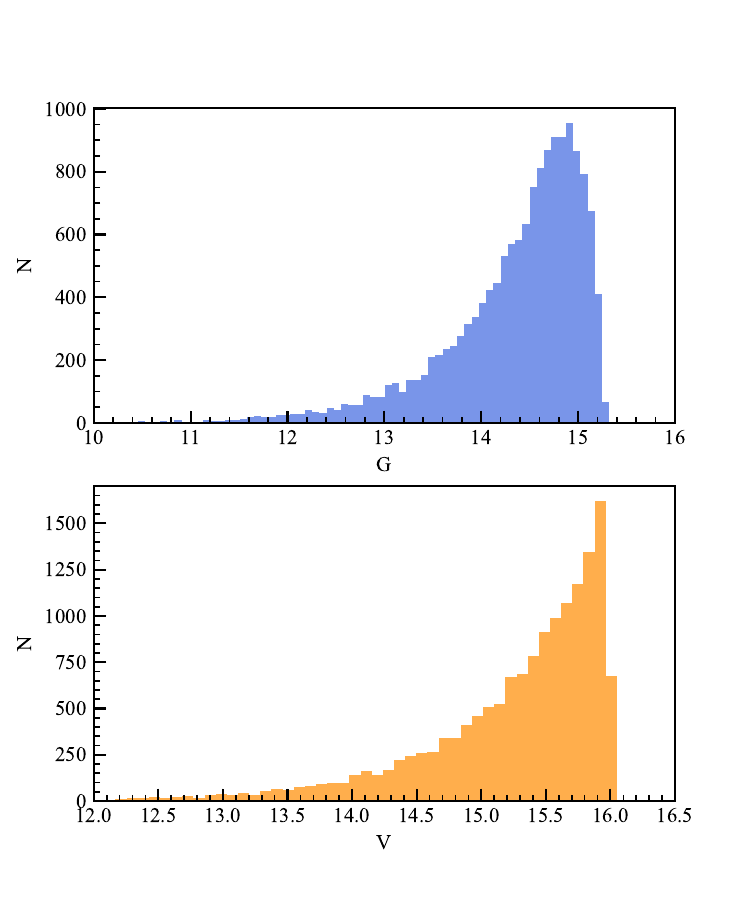}
    \caption{Distribution of apparent $G$ and $V$ magnitudes  for the P4 sample.}
    \label{fig:p4VGhist}
\end{figure}

 The mean distance of the P4 sample targets, computed from 
the distances provided by \citet{bail21},  is 
$135.4$ pc, while the median distance is $133.0$ pc. 
The  standard deviation is $\sigma_d = 59.3$ pc, while
the minimum and maximum distances observed in the sample are $d_{\text{min}} = 3.9$ pc, corresponding to the Kapteyn’s star, and 
$d_{max} = 408.8$ pc, respectively. 
The magnitude distributions in both the G and V bands are shown in Fig.\,\ref{fig:p4VGhist}.

\subsection{Validation of P4 targets and $K_S$ magnitudes}
To derive stellar radii and masses for the P4 sample targets, we need  the absolute $K_S$ magnitudes (see Sect.\,\ref{sec:radmass}) from the 2MASS catalogue \citep{skru16}.
The
2MASS near infrared colours can also be used to validate the P4 sample, since,
based on the stellar evolution model predictions, M-type stars are expected to have $J-H$, $H-K_S$ and $J-K_S$ colours in well defined ranges \citep[e.g.][]{bres12}.
For this purpose, the P4 sample was cross-matched with 2MASS using the official Gaia-2MASS cross-match procedure \citep{marr19}, ensuring completeness by systematically identifying the best 2MASS counterparts for all Gaia stars.

To validate the cross-match and ensure the reliability of the $K_S$ magnitudes for stellar parameter derivation, we considered only one-to-one matches between Gaia and 2MASS.
 In addition, in cases where multiple Gaia sources \citep[mates,][]{marr19} shared the same best 2MASS match, 
 only those mates  
 classified as M-type stars were 
 retained. 
All matched stars were compiled into a single list,
resulting in a total of 14\,909 
objects from the original 
Mdwarf sample
including 
15\,157 stars.

We note that
 no M-type stars with a 2MASS counterpart  has more than one neighbour.

\begin{figure}
   \centering
\includegraphics[width=0.5\textwidth]{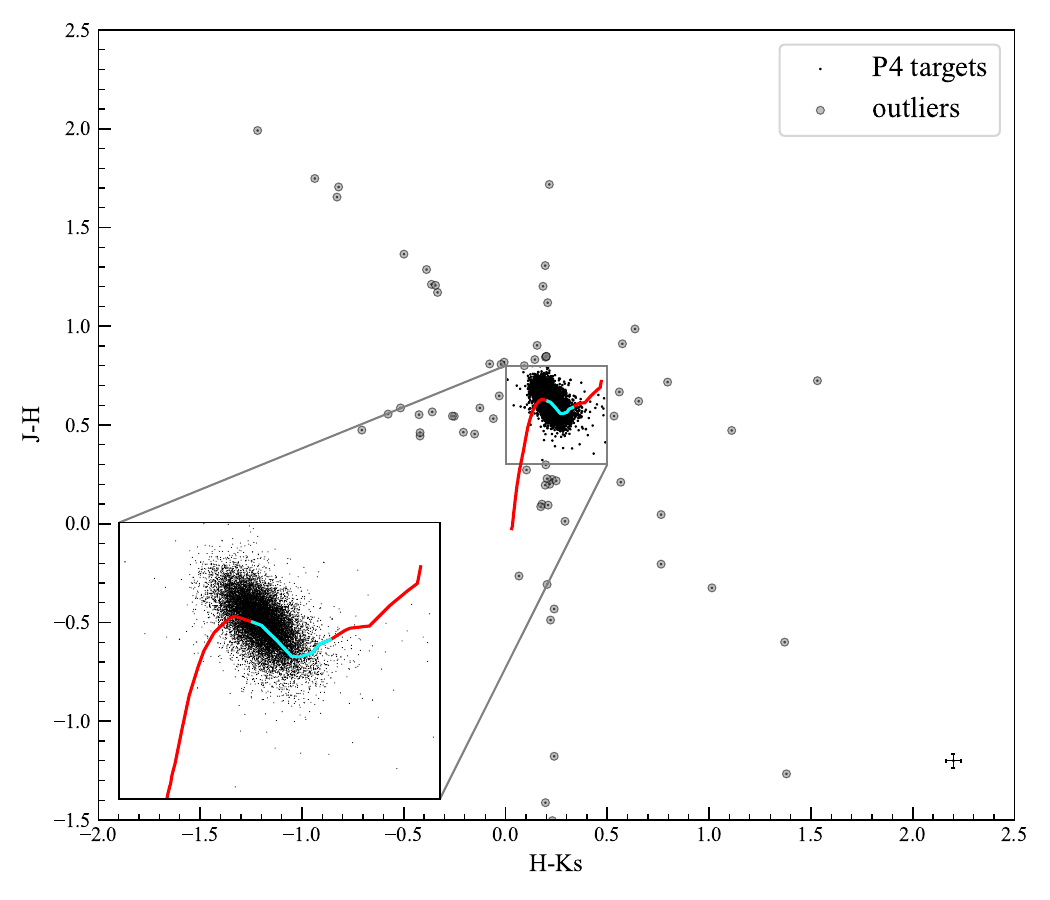}
      \caption{
Near-infrared (J–H vs. H–Ks) colour–colour diagram of the P4 sample targets. Enlarged grey circles highlight the  outliers in this  diagram. The solid red line shows the mean
colours of dwarfs from \citet{peca13}.
The cyan line indicates the subset of stars from M0V to M6V.
An inset provides a zoomed-in view of the high-density core of the distribution, allowing better visualisation of the main-sequence locus. The typical errors in the colours are indicated in the bottom-right corner.
         \label{twomassccd}}
   \end{figure}

\begin{figure}
   \centering
\includegraphics[width=0.5\textwidth]{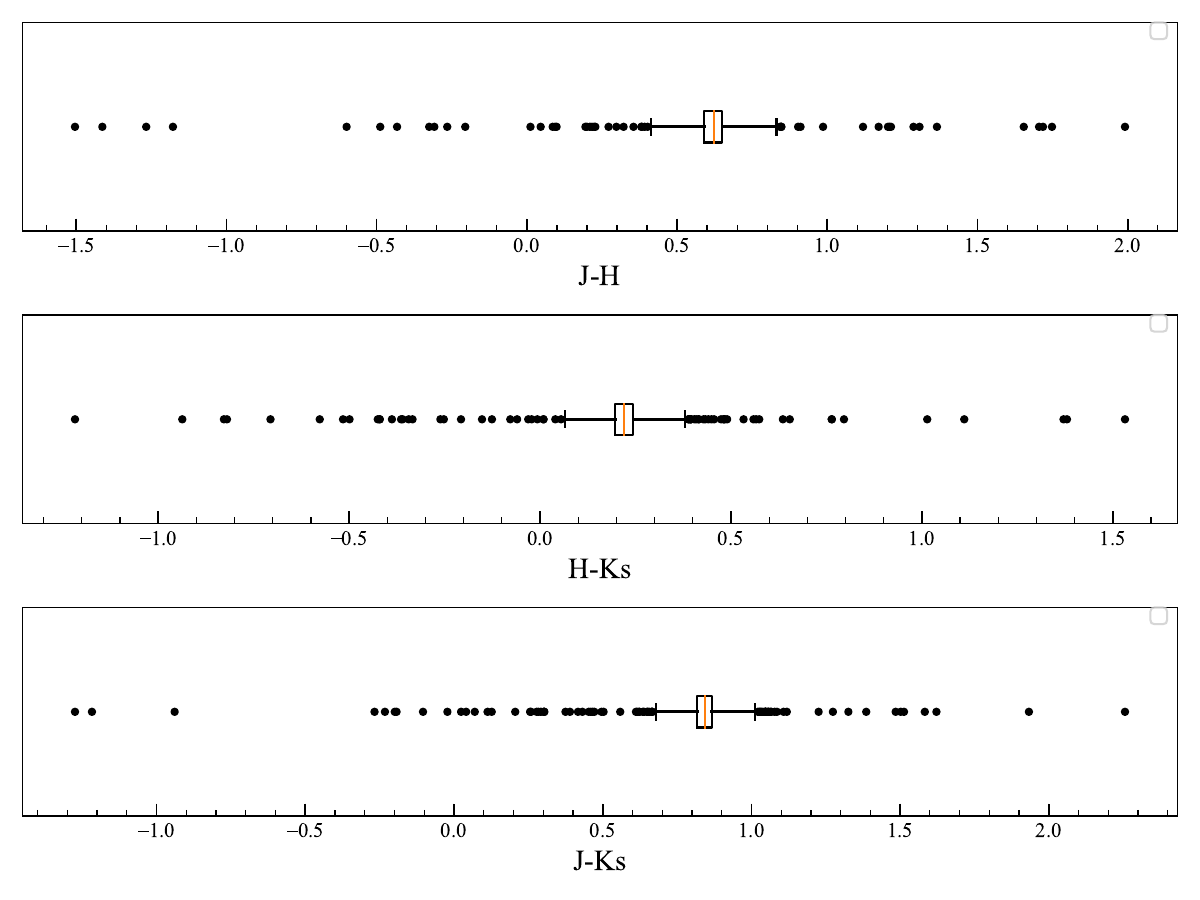}
      \caption{
      Boxplots of the $J-H$, $H-Ks$, and $J-Ks$ colour distributions for the selected P4 sample. 
Each box represents the IQR (25th to 75th percentile), with the central red line indicating the median. 
The whiskers extend to the most extreme data points that are not considered outliers. 
Individual points beyond the whiskers represent outliers, i.e., values that deviate significantly from the main distribution.}
         \label{tmassmchmad}
   \end{figure}

Figure\,\ref{twomassccd} displays the near infrared colour--colour diagram of the targets included in the P4 sample. 
The majority of the sources lie within the colour space typically occupied by M-type stars,
 indicated by the cyan curve representing the colours of stars between M0V and M6V \citep{peca13}, confirming the consistency of their photometric properties with this classification. However, a significant spread is evident, which may arise from a combination of photometric uncertainties, unresolved binarity, stellar activity (e.g., spots), and possibly metallicity variations.
In order to identify the few outliers that are not compatible with M-type stars or for which the 2MASS photometry is not reliable for 
parameter derivation, we performed a statistical analysis of the $J-H$, $H-K_S$, and $J-K_S$ colour distributions,
using the  boxplots, shown in Fig.\,\ref{tmassmchmad}. In particular, we considered the interquartile range
(IQR) i.e. IQR =Q3 - Q1, where Q3 = 75th percentile and Q1 = 25th percentile.
The median (red line), the IQR  
(black box), and the full spread of the distributions are shown in each panel. Table\,\ref{tab:color_statistics}
reports the main statistical parameters for each colour index, including the median, Median Absolute Distribution (MAD), quartiles, and the corresponding thresholds for outlier identification.

The thresholds used to identify the {\it blue outliers}, i.e. objects with the most reliable 2MASS photometry in the three bands\footnote{ph\_qual='AAA' in the JHKs bands},
that are not consistent with M-type stars,
have been defined by considering
the lower tails of these distributions, corresponding to the boundaries beyond which sources deviate significantly from the bulk of the M dwarf population.
In particular, for each of the three $J-H$, $H-K_S$ and $J-K_S$ colours,
we considered the IQR and we
defined the  {\it Outer fence} range as  [Q1–3$\times$IQR, Q3+3$\times$IQR].
We found that the colour dispersion evaluated with the MAD is very similar to the IQR, since most of the
data are concentrated around a narrow colour range.
\begin{table*}[htbp]
\caption{Statistical properties of the 2MASS color indices in the P4 sample.}
\label{tab:color_statistics}
\centering
\begin{tabular}{ccccccccc}
\hline\hline
Color & Median & MAD & Q1 & Q3 & Q1 - 3 $\times$ IQR & Q3 + 3 $\times$ IQR & Min & Max \\
\hline
J-H & 0.62 & 0.03 & 0.59 & 0.65 &  0.41 & 0.83 & -1.50 & 1.99 \\
H-Ks & 0.22 & 0.02 & 0.20 & 0.24 &  0.08 & 0.32 & -1.22 & 1.53 \\
J-Ks & 0.84 & 0.03 & 0.82 & 0.87 &  0.67 & 1.01 & -1.27 & 2.26 \\
\hline
\end{tabular}
\end{table*}

We defined \textit{blue outliers}, objects
with good 2MASS photometry, 
for which one of the three 2MASS colours is smaller than 
the lower outer fence values Q1–3$\times$IQR, corresponding to -4.72$\sigma$ (assuming that Q1 and Q3
lies at -0.675$\sigma$ and at 0.675$\sigma$ in a gaussian distribution). The criteria to select these \textit{blue outliers}  are, therefore,
$J-H<0.41$ or 
$H-K_S<0.06$ or 
$J-K_S<0.67$.

Analogously,
we defined  \textit{red outliers}, objects 
for which one of the three 2MASS colours is larger than the upper outer fence values Q3+3$\times$IQR, corresponding to 4.72$\sigma$.
The criteria to select these \textit{red outliers}  are, therefore,
$J-H>0.83$ or 
$H-K_S>0.39$ or  
$J-K_S>1.01$.

In conclusion, from the P4 sample included in the matched list, we discarded 8 stars 
with 2MASS  ph\_qual='AAA', classified as \textit{blue outliers}. We note that additionally  3 more objects with $G_{\rm BP} - G_{\rm RP} < 1.84$ were excluded, as their colours are not consistent with those of M dwarfs.
In addition, we identified 53 
\textit{red outliers} that were retained in the P4 sample as M-type targets, but for which stellar parameters were not derived, as their 2MASS colours are redder than expected for single main-sequence M-type stars, and are likely not representative of the stellar photosphere.

Based on these considerations and the analysis of the 2MASS quality flags,
the P4 targets  cross-matched with 2MASS
(one-to-one matches)
that were validated  and for which stellar parameters can be derived with the 2MASS 
Ks magnitude, are 14328 
out of 14909 (96\%).
In addition, we found 320 stars (about 2\% of the total sample)
 with one mate (a neighbour star)   for which
the 2MASS photometry is consistent with that expected for M-type stars.
These 
14\,648 (14\,328+320) 
objects can be considered valid for the P4 sample as their NIR colour-colour diagrams are  compliant with those expected for M-type stars. Stellar parameters for these targets have been derived with the 2MASS Ks magnitude.

For the remaining 261
(14\,909-14\,648)
targets, we 
concluded
that while they can be considered valid for the P4 sample, the 2MASS Ks magnitudes are not reliable for the stellar parameter derivation.

\subsection{Stellar Parameters}
\subsubsection{Effective temperatures}
In the absence of complete sets of spectroscopic data (Gaia DR3 
Radial Velocity Spectrometer
going down to $G\lesssim$14), homogeneous and consistent effective temperatures for photometrically selected samples, such as those selected for the PIC, can only be derived through the adoption of a photometric calibration.
  To this aim,  
 we adopted a set of FGK dwarfs from the dataset of \citet{casa10} and a set of M
 dwarfs from the dataset of \citet{mann15}. The FGK dwarfs were cross-matched
 with Gaia DR3, selecting
only stars within 70 pc that met our 
quality criteria,
 yielding a sample of 147 stars. 
  The qualityFlag is a bitmask used to identify issues in \textit{Gaia} astrometry and photometry, with each bit corresponding to a specific quality indicator. We defined our thresholds for these indicators based on the 95$\rm ^{th}$ percentile of their cumulative distribution for all sources with G$\leq$13. The specific indicators we flagged are: a re-normalised unit weight error (RUWE) greater than 3.6; an astrometric excess noise greater than 0.61 with a significance greater than 2; and issues related to the integrated pulse duration (IPD), including multi-peak (ipd\_frac\_multi\_peak$>14$), harmonic amplitude (ipd\_frac\_harmonic\_amplitude$>0.09$), and odd window fraction 
 (ipd\_frac\_odd\_win$>1$). Photometric quality flags include a blending fraction (beta) greater than 0.1 and a corrected BP and RP flux excess ($\lvert C^*\rvert >3\sigma_{C^*}$). Finally, we flagged sources for which specific corrections were applied, as described in \citet{riel21}---specifically, the saturation correction (Appendix C.1) and the correction from Section 8.4.
 
\begin{figure}
    \centering
\includegraphics[width=0.5\textwidth]{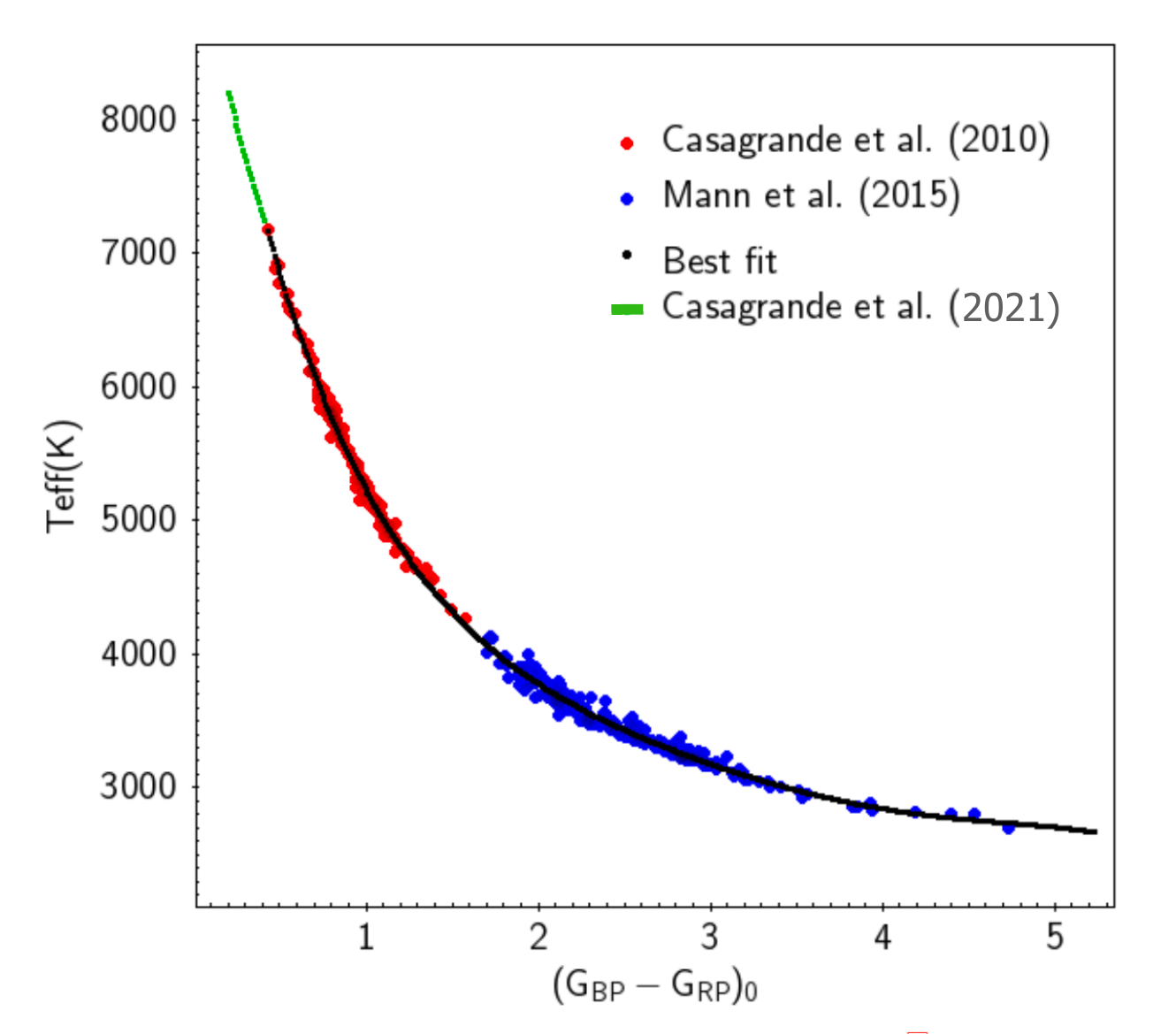}
    \caption{Best fit relationship between effective temperature and colour (Eq.\,\ref{eq:teffvscol}, black solid line). Red points represent the sample of 
    \citet{casa10}
     while blue points the sample of \citet{mann15}. The green line represents Eq. 1 of 
     \citet{casa21}
     where we imposed log g=4.438068 and [Fe/H]=0.}
    \label{fig:teffvscol}
\end{figure}

 Similarly, we considered a subset of 179 M dwarfs
 from the compilation of \citet{mann15}. We then established a polynomial
 relationship between the effective temperature ($T_{\rm eff}$) of the selected
 stars and their intrinsic 
  Gaia DR3 colour $(G_{\text{BP}} - G_{\text{RP}})_0$.
The relationship we obtained is:
\begin{equation}\label{eq:teffvscol}
  T_{\textrm{eff}} =\sum_{i=0}^5 c_i(G_{\text{BP}} - G_{\text{RP}})_0^i
\end{equation}
where the best fit coefficients are reported in Tab.\,\ref{tab:teffvscol}, and the corresponding curve is shown in Fig.\,\ref{fig:teffvscol}.
\begin{table*}[htbp]
    \centering
    \caption{Best fit coefficients of the T$_{\rm eff}$ as a function of $(G_{\text{BP}} - G_{\text{RP}})_0$ relationship represented by Eq.~\eqref{eq:teffvscol}.}
    \label{tab:teffvscol}
    \begin{tabular}{ccccccc}
        \hline
        $c_0$ & $c_1$ & $c_2$ & $c_3$ & $c_4$ & $c_5$ & $\sigma$ \\
        \hline
        9649.1817 & -7175.80969 & 3642.30312 & -1020.37499 & 146.20008 & -8.30455 & 55 K\\
        \hline
    \end{tabular}
\end{table*}

The colour range of validity is 
0.43$\leq(G_{\text{BP}} - G_{\text{RP}})_0\leq$5  and the standard deviation of the calibration is 55 K. 
To be consistent with \citet{casa10} and following the approach described in \citet{mont21}, 
 we initially calculated the effective temperature using Eq.\eqref{eq:teffvscol} assuming 
$(G_{\text{BP}} - G_{\text{RP}})_0 = (G_{\text{BP}} - G_{\text{RP}})$, i.e. assuming negligible reddening for the adopted stars.  We then estimated the monochromatic extinction from \citet{lall22}  and converted it to the extinction in all photometric bands. Then we estimated again the intrinsic colour as
$(G_{\text{BP}} - G_{\text{RP}})_0 = (G_{\text{BP}} - G_{\text{RP}})-E(G_{\text{BP}} - G_{\text{RP}})$. We iterated this procedure three times since this was sufficient to obtain convergence on the effective temperature ($\Delta{T_{\rm eff}}\leq$10 K).
\subsubsection{Radii and masses \label{sec:radmass} }
Stellar radii  for the P4 targets were derived through the empirical 
relation
$Y = a + bX + cX^2$,
determined in \citet{mann15}, 
where $X$ is the absolute Ks-band magnitude ($M_{K_S}$) and $Y$ is the stellar radius ($R_\star$).
The coefficients $a$, $b$, and $c$ 
 of the second-order polynomial fit,
 taken from \citet{mann15},
 are given in Table\,\ref{tab:tabmann}.
\begin{table*}[htbp]
    \centering
    \caption{Best fit coefficients from  \citet{mann15}}
    \label{tab:tabmann}
    \begin{tabular}{ccccccc}
        \hline
        $Y$ & $X$ & $a$ & $b$ & $c$ & $d$ & $e$ \\
        \hline
        $R_*$ & $M_{K_S}$ & 1.9515 & -0.3520 & 0.01680 & ...& ... \\
        $M_*$ & $M_{K_S}$ & 0.5858 & 0.3872 & -0.1217 & 0.0106 & -2.7262$\times10^{-4}$ \\
        \hline
    \end{tabular}
\end{table*}
This relation is based on the well-established luminosity–radius correlation for main-sequence stars, using absolute Ks magnitudes, $M_{K_S}$,
as a proxy for luminosity. In the case of M dwarfs, this correlation is particularly tight and 
shows very little dependence on 
metallicity, making $M_{K_S}$ a reliable predictor of the stellar radius.

 \begin{figure*}[tp]
    \centering
    \includegraphics[width=0.48\textwidth]{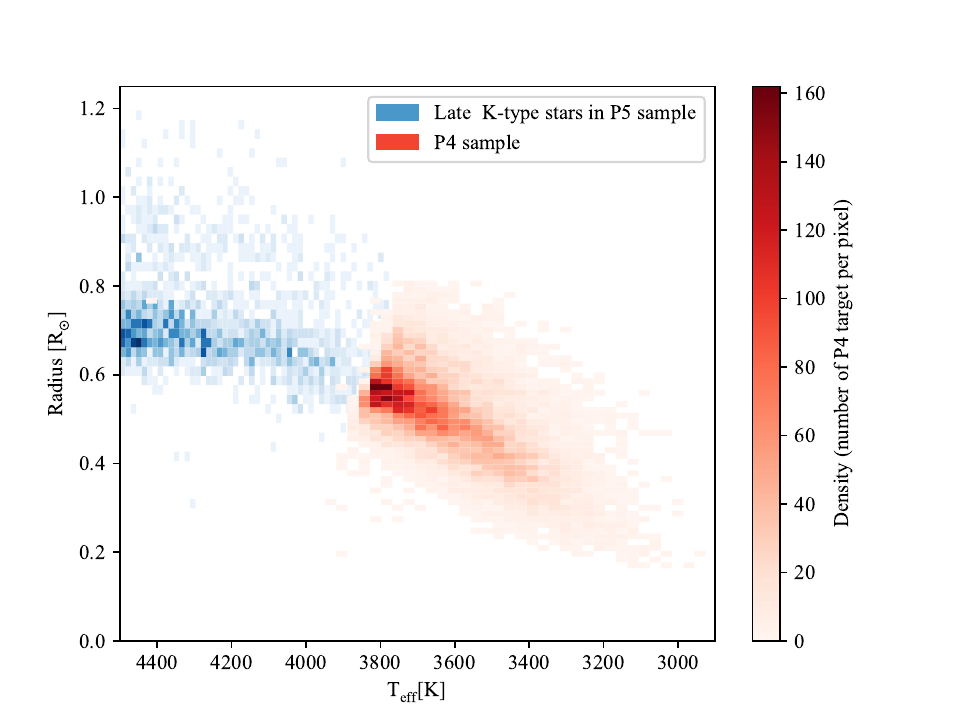}
    \includegraphics[width=0.48\textwidth]{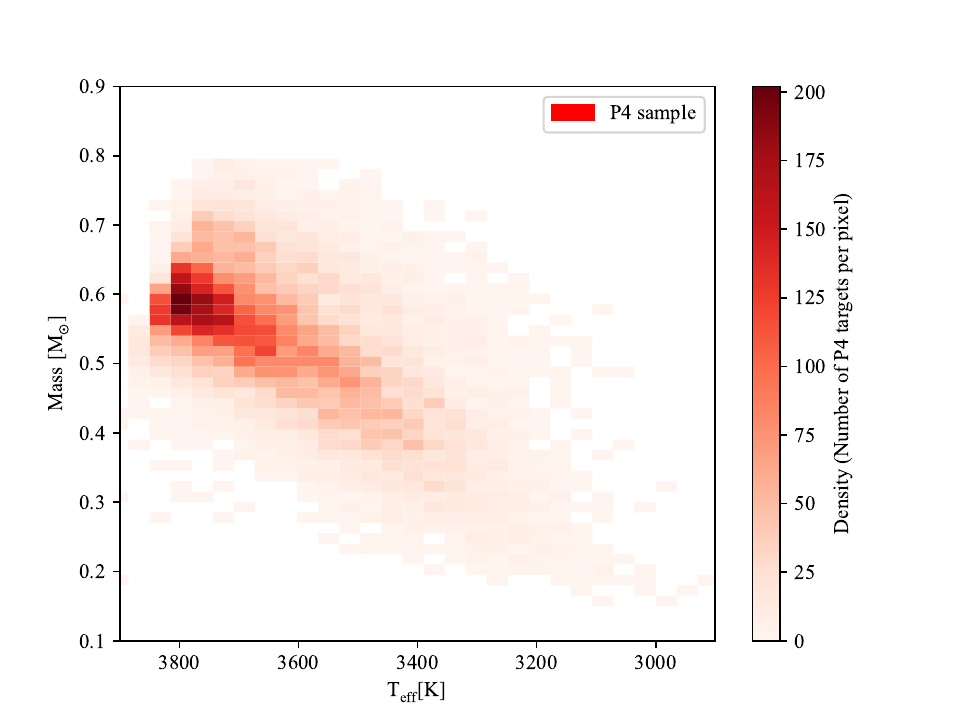}
    \caption{Distribution of stellar radii (\textit{left}) and masses (\textit{right}) as a function of effective temperature for the P4 sample (red map). The density map  in blue corresponds to late-type stars in the P5 sample. The colour intensity indicates the density of stars per bin.}
    \label{fig:mass_rad_teff}
\end{figure*}

Stellar masses  for the P4 targets have been derived through the semi-empirical relationship
$Y = a + bX + cX^2 + dX^3 + eX^4$,
where  $X$ is the absolute $K$-band magnitude ($M_{K_S}$) and $Y$ is the stellar mass ($M_\star$),
also given in \citet{mann15}.
This relation has been constructed using a fourth order polynomial fit between stellar model-derived masses and observed  absolute $K_S$
magnitudes, $M_{K_S,0}$, corrected for extinction. The coefficients of such a relation, taken from \citet{mann15}, are given in Table\,\ref{tab:tabmann}.

The validity range of these relations is
$0.1\lesssim R_*/R_\odot \lesssim 0.7$, corresponding to spectral types K7–M7, with
$4.6 < M_{K_S,0} < 9.8$,
$2700 < T_{\mathrm{eff}} < 4100$ K,
and $-0.6 < [\mathrm{Fe/H}] < 0.5$.
These limits are consistent with 
97\% of the targets in 
the P4 sample, given that 14\,162 
out of 14\,909 have $M_{K_{S,0}}$ in this range.
The number density of the stellar parameters derived for the P4 sample is shown in Fig.\,\ref{fig:mass_rad_teff}. 

In order to check the 
homogeneity of the different methods adopted to derive  the stellar radii for the P4 and P5 samples\footnote{the P5 sample includes F5-K7 type stars with $V\leq13$ \citep{mont21}},
we also show the number density
of the subsample of P5 colder than 4500\,K.
From this comparison, we conclude that the methods adopted to derive stellar radii in the P4 and P5 samples are consistent
as the distributions show a smooth continuity, indicating that no systematic discontinuities are introduced by the different methods.  Furthermore, we assume that the above relations can be extrapolated for the 474 stars with 
$3.05<M_{K_{S,0}}<4.6$
 that fall outside the validity range, as we observe no discrepancy with the radii derived for the stars
 with spectral types earlier than M.

\section{Results\label{sec:results}}
\subsection{Photometric and volume completeness}
 Based on the results presented in \citet{smar21}, we expect the completeness of our sample to be greater than 99\% for stars with $V<16$ 
 (magnitude limit imposed by the science requirement R-SCI-234), 
 as this magnitude limit is well within the range of high completeness for Gaia DR3.

To explore the 
volume completeness of our sample, we used the
\(\langle V/V_{max} \rangle\) test \citep{schm68},
originally adopted for quasars. 
By volume completeness we mean the distance (or volume) within which the P4 sample can be considered statistically complete, i.e. not affected by incompleteness due to observational limits.
This approach evaluates whether the distribution of objects in a sample remains statistically uniform within a given maximum radius \( R_{max} \), beyond which incompleteness effects may become significant.
In a uniformly distributed sample,
 \(\langle V/V_{max} \rangle\) has a Poisson distribution and its 
 expected value $E$  is 0.5,
 while  the statistical uncertainty is
$    \sigma = 1/\sqrt{12 N}$.
This uncertainty derives from the variance
\(\sigma^2 = 1/12\) of a single variable uniformly distributed between 0 and 1.
Therefore, a sample is considered complete if 
\(\langle V/V_{max} \rangle\) is statistically consistent with 0.5 within a defined confidence interval. In particular, compatibility within 
$\pm 3\sigma$
is typically adopted as a robust criterion, 
assuming purely random (Poisson) statistics.

To apply this test, we selected a subset of stars with absolute magnitudes in the range 
\( 7 \leq M_G \leq 14 \),
which is the magnitude range covered by our targets.
For each star $i$, we 
defined  the volume $V_i$, enclosed within its observed distance $d_i$.
We then defined a set of increasing values\footnote{$R_{max}=100$\,pc was chosen because Fig.\,\ref{fig:p4mg0borp0} shows that, beyond this distance, our sample no longer includes late-M stars.}
of \( R_{max} \) from 0 to 100\,pc, in steps of 1\,pc. For each \( R_{max} \), we computed the total volume
$    V_{max} = \frac{4}{3} \pi R_{max}^3$
and selected stars with \( d_i \leq R_{max} \). The mean \(\langle V/V_{max} \rangle\) value was then computed as
$    \langle V/V_{max} \rangle = \frac{1}{N} \sum_{i} \frac{V_i}{V_{max}}$,
where \( N \) is the number of stars in the sample with distances within \( R_{max} \). 

\begin{figure}
    \centering
\includegraphics[width=0.5\textwidth]{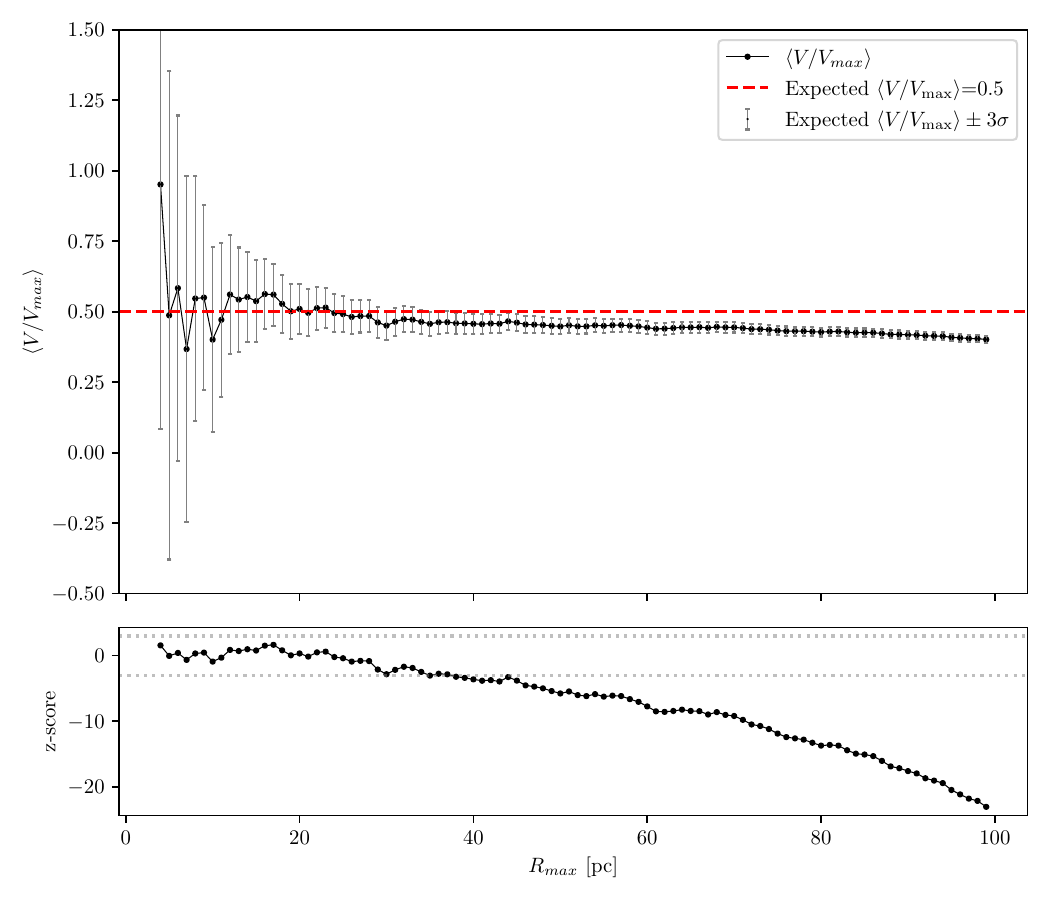}
\caption{Top panel: $\langle V/V_{\rm max} \rangle$ as a function of $R_{\rm max}$ for the P4 targets with $7 \le M_{G,0} \le 14$, where $V_{\rm max}$ is the corresponding maximum volume. The red dashed line indicates the expected value of 0.5, and the grey error bars show the 3$\sigma$ statistical uncertainty.  
Bottom panel: z-score of the deviation from the expected 0.5 value, defined as $z = (\langle V/V_{\rm max} \rangle - 0.5)/\sigma$, as a function of $R_{\rm max}$. The horizontal dotted grey lines indicate the $\pm3\sigma$ limits used to assess the completeness of the sample.}
    \label{fig:volcomp}
\end{figure}

Figure~\ref{fig:volcomp} shows \(\langle V/V_{max} \rangle\) as a function of \( R_{max} \). The horizontal dashed red line marks the expected value of 0.5 for a homogeneous distribution, while the error bars represent the \( 3\sigma \) confidence interval around this value. 
In order to establish the distance at which the P4 sample can be considered complete, we computed the deviation from 0.5 normalised by its uncertainty, i.e., the z-score defined as $z = (\langle V/V_{\rm max} \rangle - 0.5)/\sigma$. We found that the distribution remains compatible with 0.5 up to $R_{\rm max} \simeq 37$\,pc when considering a $3\sigma$ threshold.
At larger distances, deviations from the expected value indicate the onset of incompleteness due to the imposed limiting magnitude.

\begin{figure*}
    \centering
\includegraphics[width=0.9\textwidth]{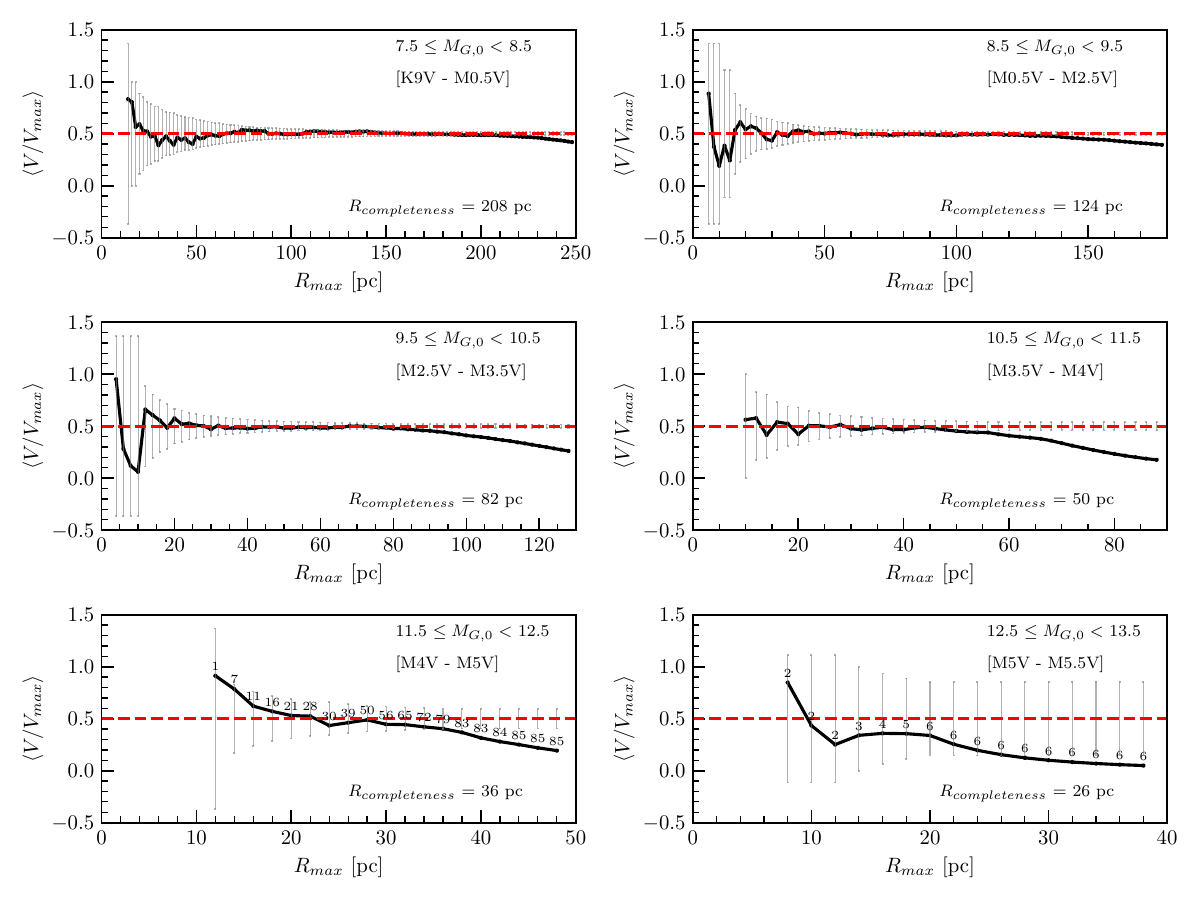}
    \caption{<$V/V_{max}$>  vs. $R_{max}$ corresponding to the volume $V_{max}$ for the P4 sample in different absolute magnitude ranges in the G band. Symbols are as in Fig.\,\ref{fig:volcomp}. The number in the bottom panels indicate the number of stars detected at each distance range.}
    \label{fig:volcomp_binned}
\end{figure*}

To evaluate the volume completeness of our sample as a function of spectral type, or, equivalently, of absolute magnitude \(M_G\), we applied the same method described above.  
Specifically, we computed the \(\langle V/V_{\mathrm{max}} \rangle\) 
statistics
in successive 1-magnitude-wide bins of \(M_G\), ranging from 
\(M_G = 7.5\) to \(M_G = 13.5\).
These bins were chosen to cover  representative ranges of late K and early to mid M-type main-sequence stars.
The results are shown in Fig.~\ref{fig:volcomp_binned}, where we report, for each magnitude bin, the completeness distance (\(R_{\mathrm{completeness}}\)), defined as the distance within which the sample is statistically consistent with a uniform distribution at the \(3\sigma\) level.  For the two magnitude bins in which the samples are more sparsely populated (bottom panels), the number of stars detected in each distance interval is also reported.

These results indicate that the volume completeness of the P4 sample declines toward later spectral types, as the corresponding stars are fainter and detectable only within shorter distances. While the P4 sample includes stars down to spectral type M5V (see Fig.\,\ref{fig:volcomp_binned}), it is overall complete only within a radius of approximately 26\,pc, confirming the previous finding from Fig.\,\ref{fig:volcomp}. At larger distances, the number of observed stars increasingly falls short of the number expected from a uniform distribution. In contrast, stars of earlier types, such as M0.5V, remain complete out to 208\,pc. 

\subsection{Galactic Spatial-Velocity Components and P4 Stellar Population Classification \label{sec:toomre}}

In this section, we 
present
the Galactic space velocities 
of the P4 targets
in order to investigate their kinematic properties and 
possible
membership to different Galactic populations.
The resulting classification gives us
an indirect estimate of the metallicity of the P4 sample by associating stars with different Galactic populations, which are known to have distinct metallicity distributions.

The Galactic spatial-velocity components $(U, V, W)$ were computed as in Maldonado et al. (in preparation). In particular, these components were derived using Gaia DR3 parallaxes, proper motions, and radial velocities, following the method outlined by \citet{mont01} and \citet{mald10}.
Radial velocities were retrieved from the Gaia Archive for 14\,625 of the 
15\,157 
stars included in the P4 sample.
The uncertainties were calculated by considering the full covariance matrix ensuring that correlations between the astrometric parameters were accounted for.

The uncertainties in the velocity components were computed using the astrometric 
parameters provided in Gaia~DR3. Since Gaia DR3 does not provide covariance 
terms for radial velocities, only the astrometric part of the covariance matrix was 
included in the error propagation. The covariance elements 
$\mathrm{cov}(\varpi,\mu_{\alpha*})$, 
$\mathrm{cov}(\varpi,\mu_\delta)$, and 
$\mathrm{cov}(\mu_{\alpha*},\mu_\delta)$ 
were obtained from the corresponding correlation coefficients 
$\rho(\varpi,\mu_{\alpha*})$, 
$\rho(\varpi,\mu_\delta)$, and 
$\rho(\mu_{\alpha*},\mu_\delta)$, provided by Gaia DR3,  
through the relation
\[
\mathrm{cov}(x,y) 
\propto \rho(x,y)\,\sigma_x\,\sigma_y,
\]
where $\sigma_x$ and $\sigma_y$ are the uncertainties of the two astrometric parameters. 
The final uncertainties in $(U,V,W)$ were then derived by propagating both the 
individual variances and the above covariance terms in quadrature.

\begin{figure}[h]
    \centering
\includegraphics[width=0.5\textwidth]{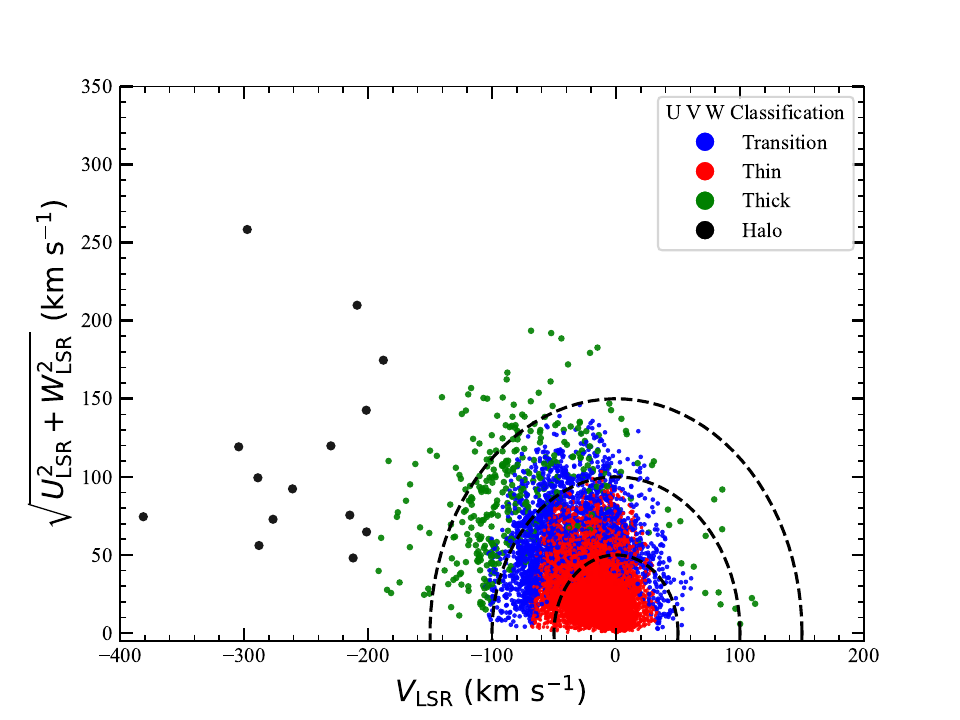}
    \caption{Toomre diagram for the P4 sample. Different Galactic populations are drawn with distinct colours. The dashed lines indicate constant total velocities, \( V_{total} = \sqrt{U_{LSR}^2 + V_{LSR}^2 + W_{LSR}^2} \), for values of 50, 100, and 150 km/s.
}
    \label{fig:toomre}
\end{figure}

The Toomre diagram for the P4 sample, shown in Fig.~\ref{fig:toomre}, provides insight into the kinematic classification of the stellar populations \citep[e.g.][]{fuhr04}.
This diagram shows the quantity \( \sqrt{U_{\mathrm{LSR}}^2 + W_{\mathrm{LSR}}^2} \), i.e. the combined radial and vertical velocity components orthogonal to the direction of Galactic rotation, as a function of \( V_{\mathrm{LSR}} \), the tangential velocity component.
Here, \( U_{\mathrm{LSR}}, V_{\mathrm{LSR}}, W_{\mathrm{LSR}} \) are the heliocentric Cartesian velocity components corrected for the solar motion with respect to the Local Standard of Rest (LSR).
The total velocity in the radial and vertical directions reflects how much a star deviates from a circular orbit in the Galactic plane and contributes to its overall kinetic energy.

In order to classify the stars into thin/thick disk populations, we use the
approach described by \citet{bens03,bens05}. In brief, by assuming that the
space velocities for the different Galactic populations have Gaussian
distributions, the authors provide equations to calculate the probabilities that a
given star belong to either the thin disk, 
the thick disk, 
or the halo. 
 For each star, the “relative probabilities” $P_{\rm thick}/P_{\rm thin}$ and $P_{\rm thick}/P_{\rm halo}$ are calculated. Thin disk stars are selected as those with $P_{\rm thick}/P_{\rm thin} < 0.1$, while thick disk stars are those with $P_{\rm thick}/P_{\rm thin} > 10$. Additionally, $P_{\rm thick}/P_{\rm halo} > 1$ is required for both thin and thick disk stars.
 We find that 11978 stars (81.9\%) belong to the 
thin disk, indicating a population with relatively low velocity dispersion and typical disk-like kinematics. A small fraction of stars (2.63\%, 385 in total) are associated with the 
thick
disk, characterised by higher velocity dispersions and older stellar ages.  
Additionally, 15.36\% of the stars (2247 in total) are classified as transition objects, as they occupy an intermediate region between the 
thin and thick
disk populations. These stars may represent a mix of both populations or stars in kinematic transition. 
Finally, a very small fraction (0.13\%, 15 in total) is identified as halo stars, which exhibit high velocities and are likely part of the Galactic halo, a population dominated by older stars on more eccentric orbits. 
This sample includes the Kapteyn’s star (GL 191, Gaia DR3 4810594479418041856) with the components ($U, V, W$) = (20.07$\pm$,0.03, 
$-$287.92,$\pm$0.10, 
$-$52.30$\pm$0.07) km\,s$^{-1}$, consistent with the values (21.1$\pm$0.3, 
$-$287.8$\pm$0.3, 
$-$52.6$\pm$0.3) km\,s$^{-1}$ given in \citet{koto05}.
\section{Discussion \label{sec:discussion}}
\subsection{Stellar radii: comparison with other studies}
\begin{figure}
    \centering
\includegraphics[width=0.55\textwidth]{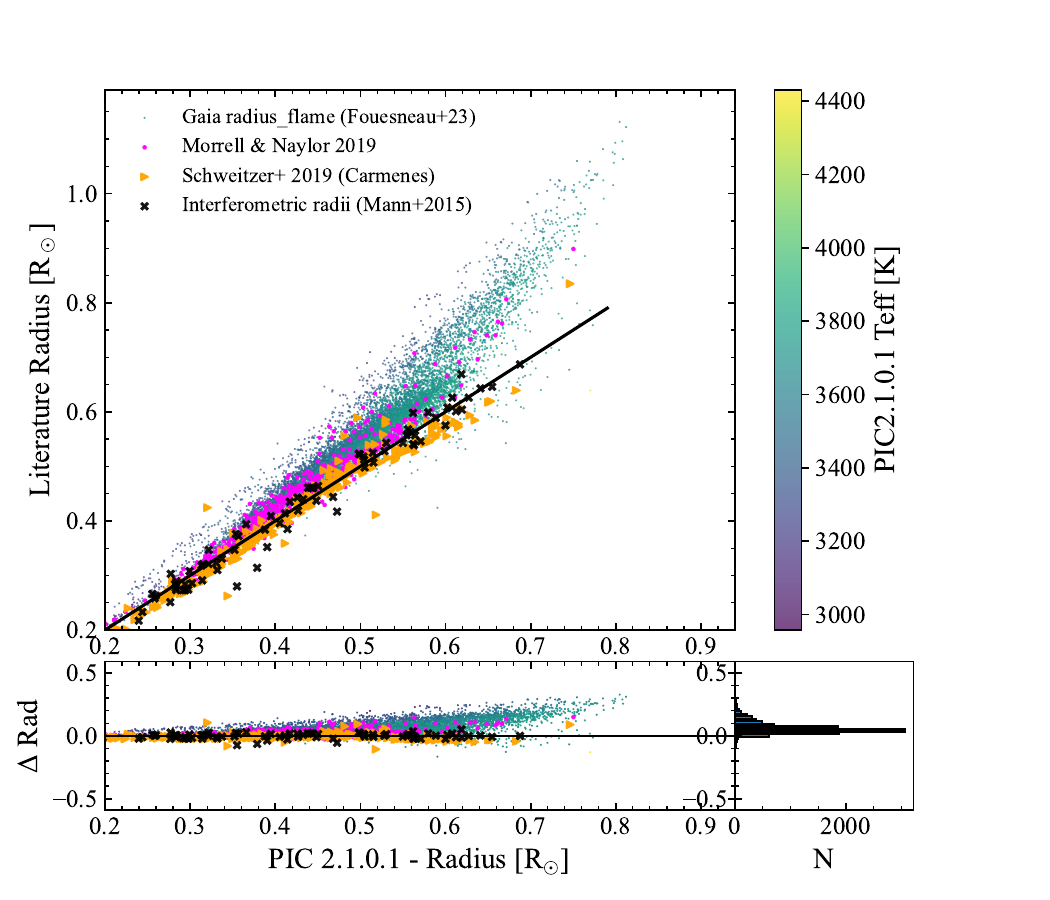}    \caption{
Comparison of stellar radii from different sources. 
\textit{Top panel}: Comparison between  
P4 radii estimated in this work (x-axis) and the 
\texttt{radius\_flame} values from Gaia DR3 (y-axis),
shown with a colour map representing the
 effective temperature. Radii from other literature works are also shown. Solid line is the 1:1 relation.
\textit{Bottom panel}: Distribution of the differences between the P4 radii and Gaia DR3 \texttt{radius\_flame} values and other external literature estimates.}
    \label{fig:radcomparison}
\end{figure}
Stellar radii included in the P4 sample have been compared with literature values.
We first considered the radii of the P4 targets released by Gaia DR3 \citep{foue23},
taken from the table \texttt{gaiadr3.astrophysical\_parameters} of the ESA Gaia archive,
namely \texttt{radius\_gspphot} and
\texttt{radius\_flame}.
The 
latter were obtained with the
Final Luminosity Age Mass Estimator 
\citep[FLAME,][]{cree23},
designed to determine stellar mass and evolutionary parameters for Gaia sources. It is based on the outputs from 
General Stellar Parametrizer from Photometry \citep[GSP-Phot,][]{andr23}
 and 
General Stellar Parametrizer from Spectroscopy
\citep[GSP-Spec,][]{reci23},
as well as astrometry, photometry, and stellar models, to estimate parameters such as radius, luminosity, gravitational redshift, mass, age, and evolutionary stage.

Results of this comparison are shown in Tab.\,\ref{tab:comparison} and Fig.\,\ref{fig:radcomparison}. 
The comparison reveals a significant discrepancy between the P4 radii and those provided by Gaia DR3, considering both \texttt{radius\_flame} and \texttt{radius\_gspphot}. In particular, the Gaia radii are systematically larger than those estimated for early type M-stars of the P4 sample, 
reaching values of up to 1\,R$_\odot$, which is surprising and unexpected for M-type stars.
  \begin{figure}
    \centering
\includegraphics[width=\columnwidth]{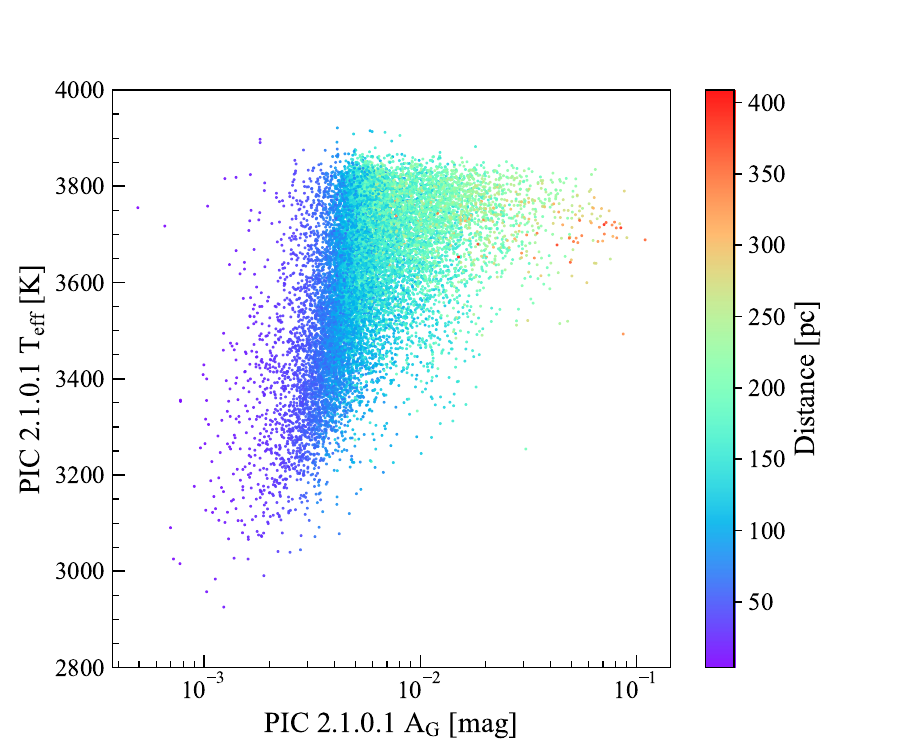}   \includegraphics[width=\columnwidth]{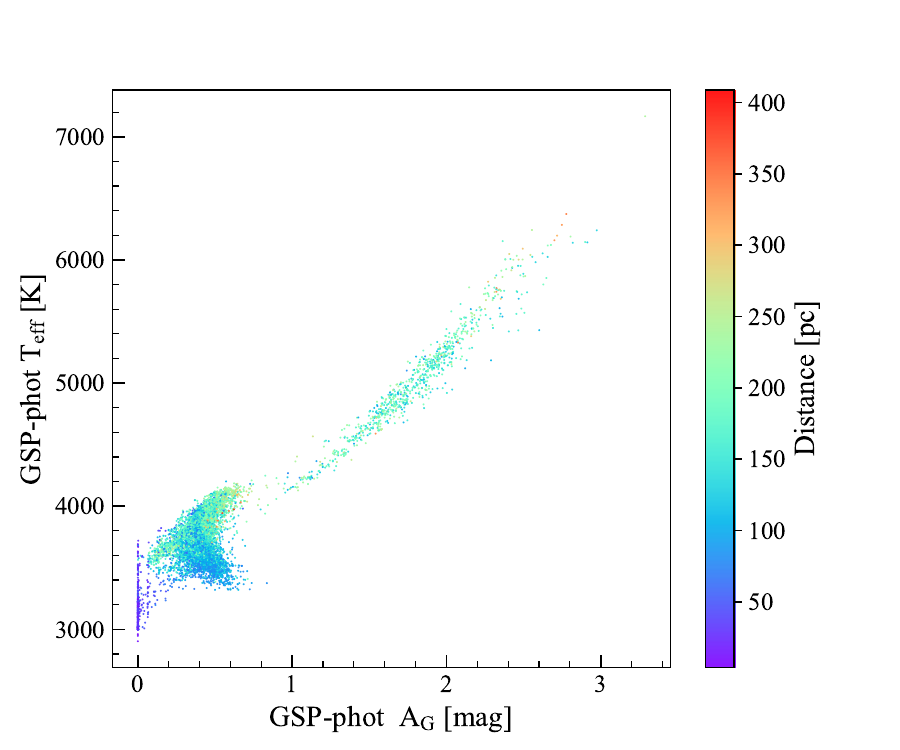}  
  \includegraphics[width=\columnwidth]{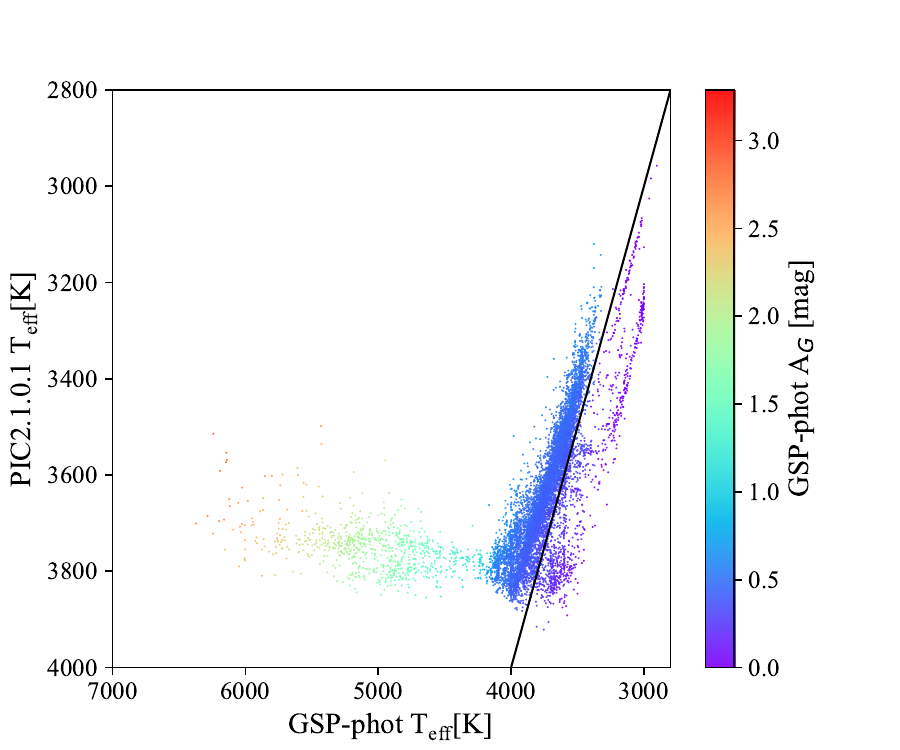} 
    \caption{
    Effective temperatures and extinction derived in PIC\,2.1.0.1 (upper panel) and Gaia DR3 (middle panel), with colour indicating the Gaia DR3 distance. 
    The bottom panel compares GSP-phot (x-axis) and PIC\,2.1.0.1 (y-axis) effective temperatures, colour-coded by GSP-phot extinction. The solid line marks the one-to-one relation.
    }
    \label{fig:aggaia}
\end{figure}
A better  agreement  is observed in the comparison with other works, focused on M-dwarfs. In particular, we considered radii obtained with two different independent methods, namely the modified SED fitting method proposed by \citet{morr19}, 
 and
 the method based on the luminosity from integrated broadband photometry, adopted in \citet{schw19}.  
For reference, the comparison with 
 the direct method of the interferometric radii by \citet{mann15}, used for the calibration adopted in this work, is also indicated.  

Table\,\ref{tab:comparison} summarises the comparison between the PIC2.1.0.1 stellar radii and values from  literature. We report the mean fractional residuals, their standard deviation, the median, the MAD, and the number of stars in common for each dataset.
Although the PIC radii show a systematic offset of about 10\% compared to Gaia DR3 values, they are generally in good agreement with other literature measurements.
As expected, for the \citet{mann15} sample (the same used to derive the calibration adopted in this work), the median fractional difference is close zero, with a MAD of 0.030, showing a negligible   systematic offset and a typical scatter of about 3\%.
The \citet{morr19} radii are slightly larger on average (median residual 0.038), while the \citet{schw19} sample shows a small negative offset (median -0.030). In all cases, the scatter is quantified using the median absolute deviation of the fractional residuals, which is lower than 3\% (if we do not consider the comparison with Gaia DR3), indicating that the PIC radii are consistent with the literature values within a few percent.

 As noted above, the \citet{morr19} radii are also slightly larger than the radii estimated for the P4 targets. This evidence is in agreement with the conclusions given by \citet{morr19} themselves, who stated that the discrepancy with interferometry-anchored methods \citep[such as][]{mann15} can arise from the inclusion of active M-dwarfs in the sample. The magnetic activity/starspot effect is known to cause larger radii for active stars compared to their inactive counterparts \citep{kima24}. Furthermore, this effect decreases toward smaller radii \citep{lope07,kima24}, and aligns  with the trend observed in our comparison.

In order to understand the disagreement with Gaia DR3 data,  we investigated other Gaia astrophysical parameters, namely \texttt{teff\_gspphot} and \texttt{ag\_gspphot}, from which the \texttt{radius\_flame} are derived using the Stefan-Boltzmann law.
Figure\,\ref{fig:aggaia} shows the comparison between the 
effective temperatures as a function
of the
extinction values $A_{G}$ 
included in the PIC\,2.1.0.1 and derived using the 
\citet{lall22} 3D interstellar extinction map
(upper panel) and those provided by
Gaia DR3 (middle panel). colour symbols represent the \citet{bail21}
distances. 
The PIC\,2.1.0.1 parameters cover the expected range, with very low extinction values   that gradually increase  with distance. No significant correlation is observed between extinction and effective temperature.
On the contrary, we find that the GSP-phot $A_{G}$ values, i.e. the extinction in the Gaia passband predicted by Gaia DR3, are unexpectedly large for nearby M-type stars, such as those in the P4 sample, which are located at close distances ($< 400$ pc) from the Sun.
In addition, we find that the GSP-phot $A_{G}$ values  are strongly correlated with the GSP-phot T$_{\rm eff}$, in particular  for the objects with
GSP-phot $A_{G} \geq 1$. 
 These sources are distributed over a wide range of distances, and their position in the 
 colour-magnitude 
 diagram deviates from a standard MS, being their effective temperatures higher than expected.
We discard the hypothesis that the subgroup with GSP-Phot $A_G>1$ are actually hotter stars reddened by dust located behind nearby star forming regions, since their spatial distribution across the PLATO field is uniform, similarly to the rest of the sample.
 The bottom panel of Fig.\,\ref{fig:aggaia} highlights this behaviour by directly comparing the effective temperatures derived from GSP-phot and PIC\,2.1.0.1. A systematic deviation from the one-to-one relation is visible, especially for stars with GSP-phot extinction values $A_{G}>1$, for which GSP-phot T$_{\rm eff}$	
  are significantly overestimated with respect to PIC\,2.1.0.1.
A consistent fraction of the population shows  GSP-phot temperatures in the range \( 3500 \lesssim T_{\rm eff} \lesssim 4200\,\mathrm{K} \)
and moderate extinction (\( 0.2 \lesssim A_{G} \lesssim 0.6\ \)), which are
significantly higher than those 
tabulated
in the PIC\,2.1.0.1. 
Similar discrepancies in GSP-phot A$_G$ and  GSP-phot T$_{\rm eff}$
were already highlighted  by \citet{cree23b}, for stars with T$_{\rm eff}<4500$\,K, 
in their comparison of high-quality Gaia DR3 astrophysical parameters with reference values from isochrones.
A possible reason for this discrepancy is the well-known temperature-extinction degeneracy which has  been observed in Gaia DR3 data for cool stars, as discussed in \citet{andr23}. 
\begin{table*}[htbp]
\centering
\caption{Literature radii comparison. Statistical values for fractional residuals, defined as $\frac{R_{\text{lit}} - R_{\text{PIC}}}{R_{\text{PIC}}}$, where $R_{\text{PIC}}$ is the PIC2.1.0.1 radius.}
\label{tab:comparison}
\begin{tabular}{cccccc}
\hline
Ref. & $\text{Mean}(\frac{\Delta R}{R_{\text{PIC}}})$ & $\sigma_{\Delta R}/R_{\text{PIC}}$ & $\text{Median}(\frac{\Delta R}{R_{\text{PIC}}})$ & $\text{MAD}(\frac{\Delta R}{R_{\text{PIC}}})$ & N \\
\hline
Gaia DR3 radius\_flame & 0.112 & 0.071 & 0.099 & 0.035 & 9909 \\
Morrell \& Naylor 2019 & 0.043 & 0.047 & 0.038 & 0.024 & 606 \\
Schweitzer+ 2019 & -0.027 & 0.050 & -0.030 & 0.026 & 291 \\
Mann+15 & -0.008 & 0.053 & -0.001 & 0.030 & 83 \\
\hline
\end{tabular}
\end{table*}

Our results are in agreement with the recent findings of \citet{kima24}, who presented a model-independent approach based on the surface brightness-colour relation (SBCR) and Gaia DR3 colours, calibrated using interferometric angular diameters. \citet{kima24} found a clear systematic trend between 
their results and analogous 
SBCR radii from \citet{sals21} based on G-Ks colours, for stars with $M < 0.6 M_\odot$. On the contrary, a  good agreement was observed with \citet{mann15}, on which the calibration for deriving the radii of the P4 sample stars is based. 

A detailed comparison by \citet{kima24} of SBCR radii with those derived from Gaia DR3 GSP-Phot and FLAME shows that, for M-type stars (radii $<$0.6,R$_\odot$), an 
offset between 10\% and 25\% is observed,
with Gaia DR3 GSP-Phot and FLAME radii tending to be systematically larger than those inferred from SBCR,
depending on the models adopted for the GSP-Phot or FLAME radius estimates. They also show that a comparison between radii derived from direct angular diameter measurements, used for  their calibration sample, and GSP-Phot radii reveals a systematic 
offset of about 10\%,
with radii derived from direct angular diameter measurements resulting smaller than the Gaia GSP-Phot ones.

Based on these considerations, we conclude that the Gaia DR3 FLAME and GSP-Phot parameters, including stellar radii, effective temperatures, and $A_{G}$, should be used with caution for stars with $T_{\rm eff} <$ 4200 K, as evidence of systematic overestimation of these parameters has been found both in our analysis and in previous studies.

\subsection{Luminosity spread in low-mass stars: effects of metallicity and magnetic activity}
One of the most intriguing unresolved issues concerning M-type stars in the Gaia era is the observed spread in the CAMD. While M-type stars in open clusters generally follow a well-defined sequence, field stars exhibit a broader distribution, as seen in Fig. \ref{fig:p4mg0borp0}. This spread suggests the presence of additional physical effects influencing the luminosity of M dwarfs beyond the expected evolutionary trends.

A possible explanation for this broadening is the combined effect of metallicity and magnetic activity. Metallicity variations and radius inflation can shift the position of stars in the CAMD by altering their intrinsic luminosities and colours, as metal-rich stars tend to be redder and fainter for a given mass. 

\begin{figure}
    \centering
\includegraphics[width=0.5\textwidth]
{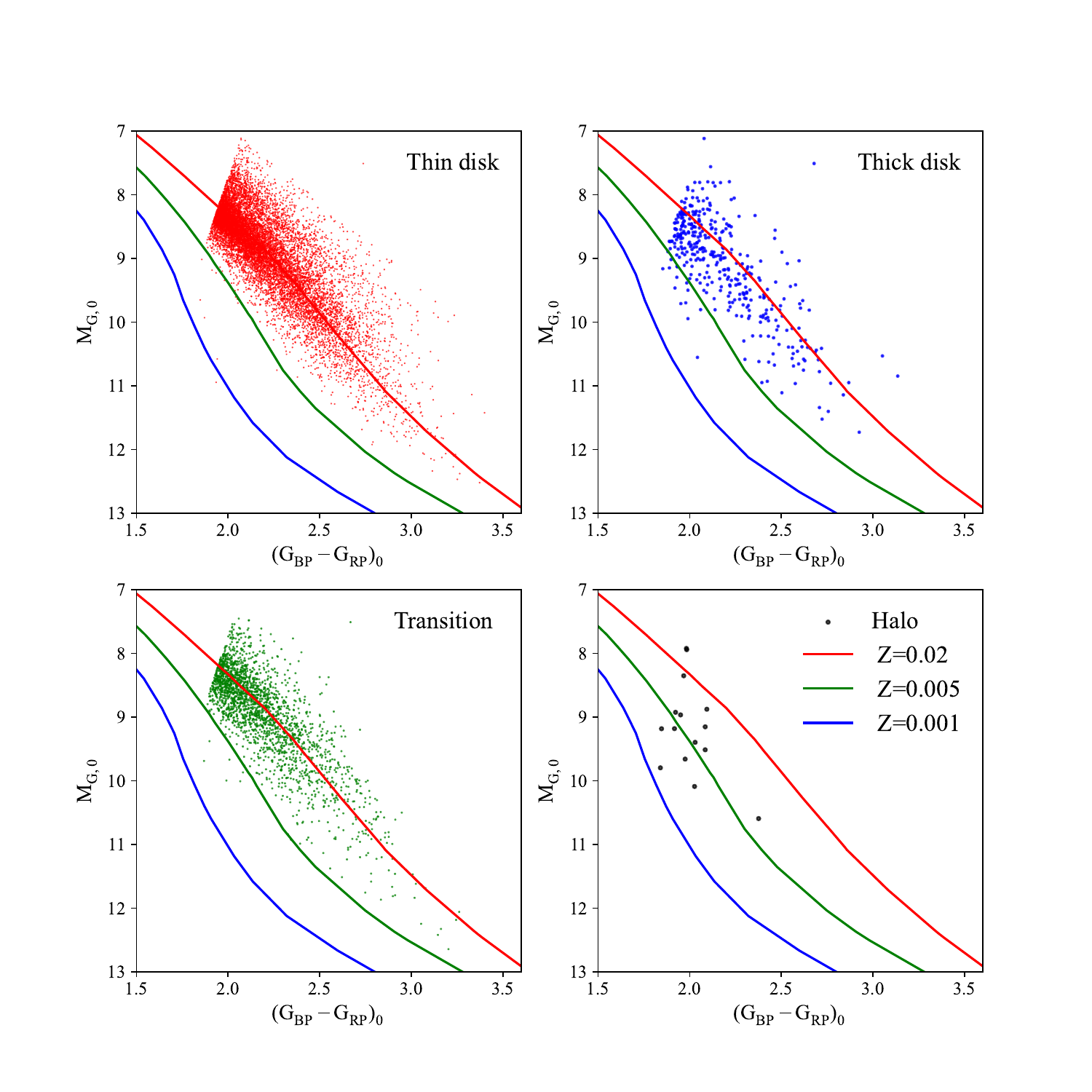}    \caption{CAMD of the different P4 galactic populations, compared with 1 Gyr isocrones computed at three different metallicity  values \citep{bres12}.}
    \label{fig:spreadfehgalcomp}
\end{figure}

In order to highlight these effects we considered
the classification based on the probability of the stars of being 
thin, thick, halo or transition stars
derived as described in Sect.\,\ref{sec:toomre}.
The four samples of targets are shown in Fig.\,\ref{fig:spreadfehgalcomp}.
As already noted, the vast majority of P4 targets
are 
thin
disk stars, consistent with a relatively young population, while only a small fraction of P4 targets are classified as 
thick disk and halo 
stars,
consistent with 
an older
population,
expected to be more metal-poorer than the 
thin
disk stars. 

To statistically compare the   different Galactic components, we performed the two-sample Kolmogorov-Smirnov (KS) test \citep{mass51} and the Anderson-Darling (AD) test \citep{ande52}. 
The KS and AD tests assess whether two samples come from the same underlying distribution, with the AD test being more robust and sensitive 
 to differences in the tails of the distributions.

\begin{figure}
    \centering
\includegraphics[width=0.45\textwidth]{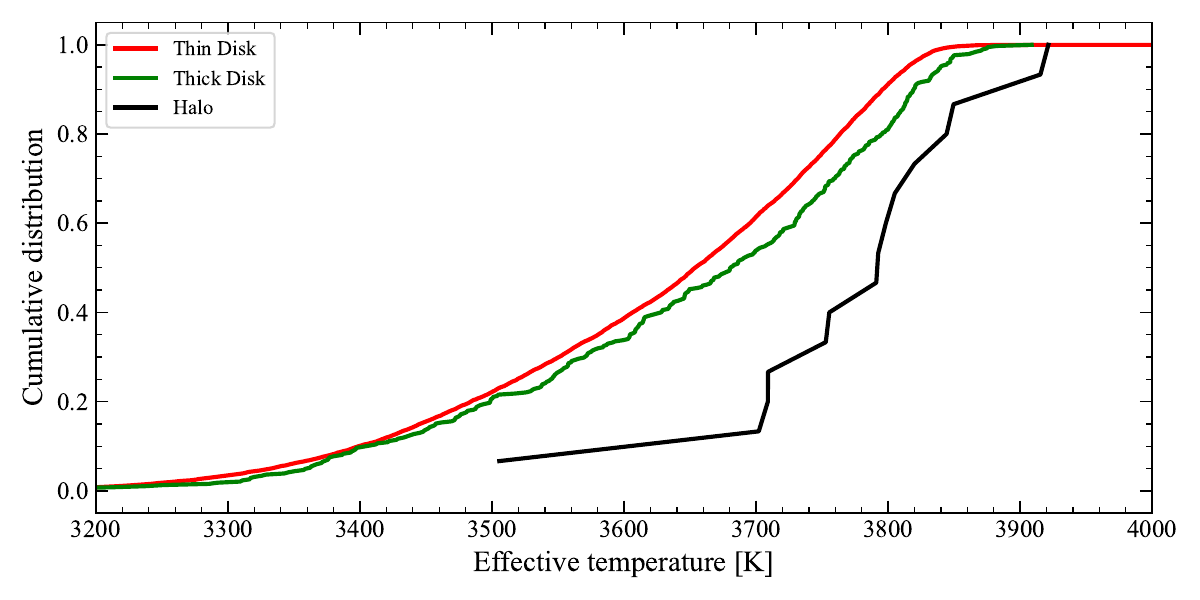} 
\includegraphics[width=0.45\textwidth]{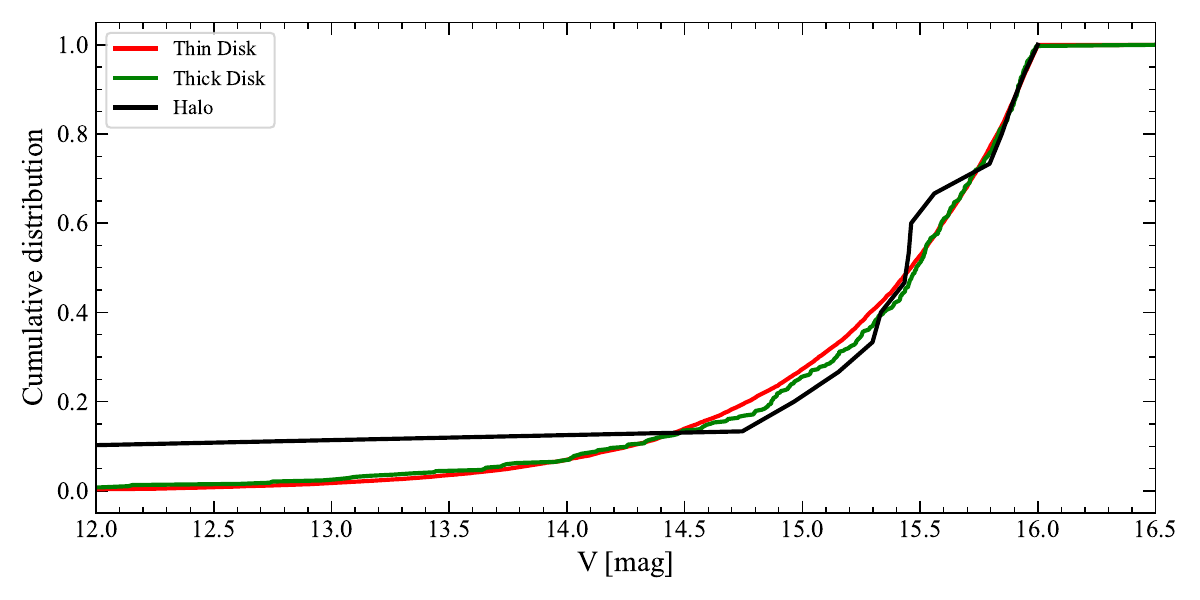} 
\includegraphics[width=0.45\textwidth]{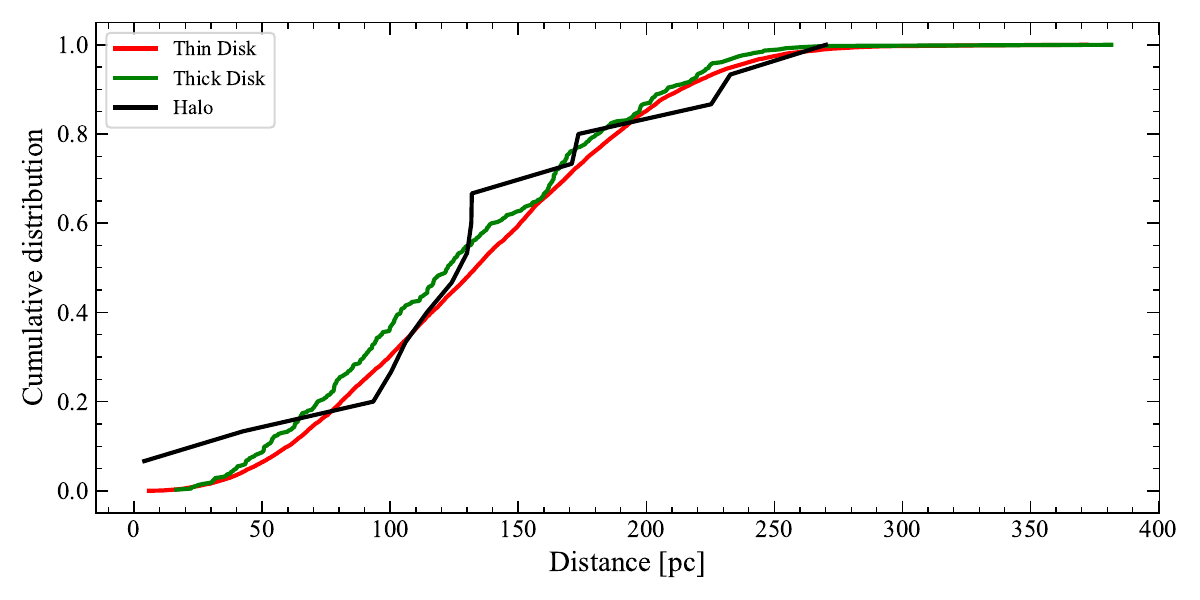} 
\caption{Cumulative distributions of effective temperature (top panel), $V$ magnitude (middle panel) and distance (bottom panel)  of the three different Galactic populations.}
    \label{fig:kstest}
\end{figure}

We applied the KS and AD tests to the effective temperature distributions of stars in the three different Galactic populations: 
thin disk, thick disk, and halo.
The cumulative distributions are shown in Fig.\,\ref{fig:kstest}, while the results of the tests are given in Table\,\ref{tab:galcomp_stattest}. We did not 
consider
the 
transition
sample, since it includes unclassifiable targets. 
\begin{table*}[htbp]
\centering
\caption{Results of the KS and AD statistical tests for comparisons between the distributions of effective temperature, distance, and apparent $V$ magnitude among the three Galactic populations. \label{tab:galcomp_stattest}}
\begin{tabular}{llcccc}
\hline\hline
 & \textbf{Populations} & \textbf{KS Stat.} & \textbf{KS p-val.} & \textbf{AD Stat.} & \textbf{AD p-val.} \\
\hline
\multirow{3}{*}{\textbf{Effective Temperatures}} 
    & Thin disk vs thick disk & 0.1041 & $5.65 \times 10^{-4}$  & 13.7656 & $< 0.001$ \\
    & Thick disk vs halo      & 0.4762 & $1.60 \times 10^{-3}$ & 7.0925  & $< 0.001$ \\
    & Thin disk vs halo       & 0.5536 & $7.51 \times 10^{-5}$ & 15.4174 & $< 0.001$ \\
\hline
\multirow{3}{*}{\textbf{V Magnitude}} 
    & Thin disk vs thick disk & 0.0439 & 0.456   & $-0.3552$ & 0.250 \\
    & Thick disk vs halo& 0.1273 & 0.950   & $-0.5730$ & 0.250 \\
    & Thin disk vs halo       & 0.1376 & 0.903   & $-0.2058$ & 0.250 \\
\hline
\multirow{3}{*}{\textbf{Distance}} 
    & Thin disk vs thick disk & 0.0786 & 0.0188  & 3.7200  & 0.0103 \\
    & Thick disk vs halo      & 0.1939 & 0.583   & 0.0062  & 0.250 \\
    & Thin disk vs halo       & 0.1763 & 0.677   & $-0.0946$ & 0.250 \\
\hline
\end{tabular}
\end{table*}

Since all KS p-values are below the conventional 0.05 threshold, we reject the null hypothesis that the compared samples are drawn from the same parent distribution,
in all cases. Similarly, the AD p-values are all $\leq 0.001$, further confirming significant differences in the effective temperature distributions among the Galactic populations. As expected, the largest discrepancy is found between the 
thin disk and the halo 
(AD statistic = 15.4), indicating a stronger divergence than that between the 
thick disk and the halo 
(AD statistic = 7.1), which is nonetheless also statistically significant. These results support the scenario in which stars from different Galactic components originated from distinct formation and evolutionary processes.
Consequently, they suggest that part of the observed spread could be attributed to the inclusion of stars with varying metallicity,
as we can see in  Fig.\,\ref{fig:spreadfehgalcomp}, where PARSEC isochrones at 1 Gyr and different metallicities (Z=0.02, 0.005, 0.001) are overplotted. 

The figure highlights how the location of the few halo stars is consistent with the low-metallicity isochrones, while higher-metallicity models are consistent with thin disk stars. 
The thick disk population, as expected, lies in between, following an intermediate metallicity sequence.
This supports the interpretation that the spread in the colour–magnitude distribution is at least partly driven by metallicity differences across the Galactic components.

To rule out the possibility that our results are influenced by selection effects, we also compared the cumulative distributions of the $V$ magnitudes, used in the selection of the P4 sample, and the distances. As shown in Fig.\,\ref{fig:kstest}, these distributions largely overlap, suggesting that the three Galactic populations share similar characteristics in these parameters.

This visual impression is supported by the KS and AD statistical tests, which quantify the significance of any differences between the distributions. The results, summarised in Table\,\ref{tab:galcomp_stattest}, show that for both $V$ magnitude and distance, the p-values are well above the conventional threshold of 0.05, indicating no statistically significant differences. Thus, the distributions are statistically compatible.

The only statistically significant result is the difference between the distance distribution of the 
thin disk population and that of the 
thick disk
population.
However, we consider this 
fact
unlikely to introduce a relevant selection bias in our analysis.
We conclude that
the observed differences in the effective temperature distributions can be confidently attributed to metallicity.

\begin{figure}
    \centering
\includegraphics[width=0.5\textwidth]{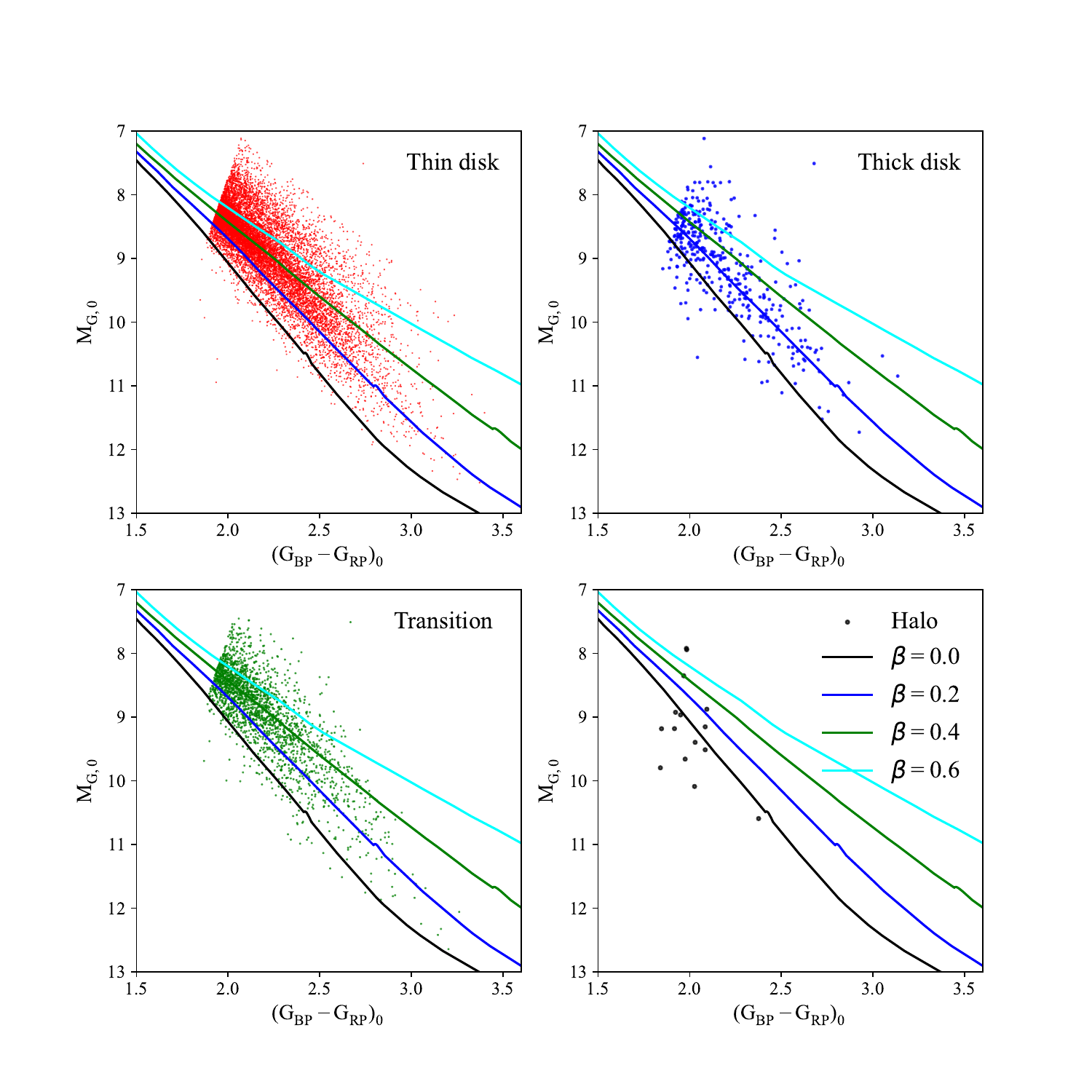}    \caption{CAMD of the different P4 galactic populations, compared with a 200 Myr solar metallicity isocrone with different $\beta_{spot}$  values \citep{fran22}.}
    \label{fig:spreadgalcomp}
\end{figure}

On the other hand, the magnetic activity, particularly in low-mass stars, can lead to radius inflation and starspot coverage, affecting the observed magnitudes and colours. For example,  \citet{some17,jeff21,fran22} 
have shown that starspots can have a significant impact on age determinations. 
Therefore, these effects should be carefully disentangled to properly interpret the distribution of M dwarfs in the CAMD and to understand the underlying stellar physics at play.
To investigate these possible activity effects,  we compared in Fig.\,\ref{fig:spreadgalcomp}
the position in the CAMD of the stars in the different 
Galactic populations with a solar metallicity 200\,Myr old isochrone,  calculated with the Pisa evolutionary code including the effect of starspots \citep{togn11,dello12,togn15,togn15a,togn18,fran22}. 
Such effect is quantified through the 
parameter 
\begin{equation}
  \beta=\frac{A_{\rm spotted}}{A_\star}\left(1-\frac{T^4_{\rm spot}}{T_0^4}\right),   
\end{equation}
that represents the effective spot coverage fraction, accounting for both the ratio between the spotted and unspotted stellar surface temperatures,  ${\rm T}_{\rm spot}/{\rm T}_{\rm 0}$  and the actual fraction of the surface covered by spots.
The parameter $\beta$ provides a direct link between the unspotted stellar effective temperature $T_0$ predicted by stellar models and the observed effective temperature, which is reduced by the presence of spots, according to
\begin{equation}
   T_{\rm eff} = T_0 \,(1-\beta)^{1/4}.
\end{equation}

For a given  isochrone with age $<200$\,Myr,  the inclusion of the starspot effects is parametrised by the four $\beta_{\rm spot}$ = 0.0, 0.2, 0.4 and 0.6 values, with $\beta_{\rm spot}$=0.0 corresponding to unspotted isochrones (see Fig.\,\ref{fig:spreadgalcomp}).  

Since our samples include stars of any age, such models suffer from the $\beta$-age degeneracy that hampers a
reliable estimate of stellar ages and simultaneously constrain the most suitable $\beta$ value. However, 
the comparison that we present shows that, in case of young stars ($\lesssim200$\,Myr),  even starspots effects can contribute to the observed spread.
In fact, this effect results in a reduced effective temperature that can enhance the spread in the CAMD for a given age.

Additional factors such as unresolved binarity, potential uncertainties in distances, or reddening effects may also contribute to the observed spread, which is not limited to M dwarfs but is also seen in earlier-type stars. Taken together with metallicity and activity, these considerations suggest that the origin of the spread remains complex and yet lacks a conclusive explanation.

\section{Conclusions \label{sec:conclusions}}
In this work, we presented the selection, validation process and properties  of the P4 sample
included in  the PLATO Input Catalogue
and
falling in the LOPS2 field.

The selection criteria for the P4 sample were designed to ensure compliance with PLATO's scientific requirements. The selection process relied on an updated photometric calibration to convert Gaia magnitudes into the Johnson-Cousins system, as well as a rigorous definition of magnitude and colour boundaries in the CAMD.
A new photometric calibration has been used to derive effective temperatures. The final P4 sample consists of 
15\,157 
targets with a mean distance of 135.4 pc and a distribution extending up to approximately 409 pc.
We validated the P4 targets by cross-matching with 2MASS to obtain $K_S$ magnitudes, essential for deriving stellar radii and masses using the \citet{mann15} photometric calibration.

The completeness analysis confirms that the P4 sample is nearly complete ($>99\%$) for stars with $V<16$, ensuring a reliable selection within the PLATO magnitude range. 
The 
$\langle V/V_{\rm max} \rangle$
test confirms that our sample is statistically uniform and volume-complete up to $\sim$26 pc. For larger distances, completeness depends strongly on spectral type: the sample remains complete up to $\simeq$50 pc for stars earlier than M4V, while at $\simeq$200 pc, only the earliest M types (M0V–M1V) are still fully represented. This highlights a progressive decline in completeness with later spectral subtypes and increasing distance.

Our comparison with the literature demonstrates that the radii we estimated for M-type stars are in agreement with those obtained with two independent radius estimation methods, all of which take into account  the intrinsic complexity of M stars. In contrast, our radii disagree with those derived from Gaia DR3 data, as found in \citet{kima24}.  

Finally, we conclude that the observed spread in the CAMD may be attributed to both metallicity effects and magnetic activity. To estimate the metallicity effects, we classified the P4 targets in 
different populations, determined using the Galactic spatial-velocity components 
$(U, V, W)$. Their distribution in the CAMD suggests that metallicity plays a role in the observed spread, even though it is 
evident  only for the few targets classified as halo
stars, which include metal-poor stars. This is expected, given that our sample is dominated by 
thin disk
stars, for which a near-solar metallicity is expected. In addition, the spread is also likely 
affected
by magnetic activity, which leads to a reduction in both effective temperature and luminosity, thereby significantly enhancing the observed dispersion.
Other factors, such as unresolved binarity, uncertainties in distances, and reddening, may also contribute, and the overall origin of the spread remains complex and not yet fully understood.

\begin{acknowledgements}
This work presents results from the European Space Agency (ESA) space mission PLATO. The PLATO payload, the PLATO Ground Segment and PLATO data processing are joint developments of ESA and the PLATO mission consortium (PMC). Funding for the PMC is provided at national levels, in particular by countries participating in the PLATO Multilateral Agreement (Austria, Belgium, Czech Republic, Denmark, France, Germany, Italy, Netherlands, Portugal, Spain, Sweden, Switzerland, Norway, and United Kingdom) and institutions from Brazil. Members of the PLATO Consortium can be found at \url{https://platomission.com/}. The ESA PLATO mission website is \url{https://www.cosmos.esa.int/plato}. We thank the teams working for PLATO for all their work.
VN, GP, MM, SB, SD, VG, DM, LM, IP, LP, RR acknowledge support from PLATO ASI-INAF agreements n. 2022-28-HH.0.\\
    This work has made use of data from the European Space Agency (ESA) mission
{\it Gaia} (\url{https://www.cosmos.esa.int/gaia}), processed by the {\it Gaia}
Data Processing and Analysis Consortium (DPAC,
\url{https://www.cosmos.esa.int/web/gaia/dpac/consortium}). Funding for the DPAC
has been provided by national institutions, in particular the institutions
participating in the {\it Gaia} Multilateral Agreement. PMM, SM, GA, MF acknowledge financial support from the ASI-INAF agreement n. 2022-14-HH.0.
L.P, J.M., L.A., C..A, \& E.F. acknowledge support from the Italian Ministero dell'Università e della Ricerca and
the European Union - Next Generation EU through project PRIN 2022 PM4JLH ``Know your little neighbours: characterising low-mass stars and planets in the Solar neighbourhood''.
SC acknowledges financial support from PRIN-MIUR-22: CHRONOS: adjusting the clock(s) to
unveil the CHRONO-chemo-dynamical Structure of the Galaxy” (PI: S. Cassisi)
funded by the European Union - Next Generation EU, and Theory grant INAF
2023 (PI: S. Cassisi). U.H. acknowledges support from the Swedish National Space Agency
(SNSA/Rymdstyrelsen).
We would like to thank B. Rojas-Ayala, N. Nardetto, M. Schlecker, D. Stamatelos, and N. Miller for their useful comments and suggestions. We thank the anonymous referee for the careful reading of the manuscript and for the valuable comments and suggestions, which helped improve the clarity and the quality of this work.

\end{acknowledgements}
\bibliography{bibdesk}{}
\bibliographystyle{aa}
\end{document}